\documentclass[preprintnumbers,showpacs,amsmath,amssymb,floatfix,9pt,prd,onecolumn,
superscriptaddress,nofootinbib]{revtex4}
\usepackage{latexsym}
\usepackage{epsfig}
\usepackage{amssymb}
\begin{document}

\title{Physical Characteristics and Maximum Allowable Mass of Hybrid Star in the Context of $f(Q)$ Gravity }

\author{Piyali Bhar}
\email{piyalibhar90@gmail.com}
\affiliation{Department of Mathematics, Government General Degree College Singur, Hooghly,
 West Bengal 712409, India}

 \author{Sneha Pradhan}
\email{snehapradhan2211@gmail.com}
\affiliation{Department of Mathematics, Birla Institute of Technology and Science-Pilani, Hyderabad Campus, Hyderabad-500078, India}

\author{Adnan Malik}
\email{adnan.malik@zjnu.edu.cn: adnan.malik@skt.umt.edu.pk}
\affiliation{School of Mathematical Sciences, Zhejiang Normal University, Jinhua, Zhejiang, China.}

\affiliation {Department of Mathematics, University of Management and Technology, Sialkot Campus, Pakistan}

\author{P.K. Sahoo}
\email{pksahoo@hyderabad.bits-pilani.ac.in}
\affiliation{Department of Mathematics, Birla Institute of Technology and Science-Pilani, Hyderabad Campus, Hyderabad-500078, India}
\affiliation{Faculty of Mathematics \& Computer Science, Transilvania University of Brasov, Eroilor 29, Brasov, Romania}

\begin{abstract}
\begin{center}
\textbf{Abstract}\\
\end{center}

In this study, we explore several new characteristics of a static anisotropic hybrid star with strange quark matter (SQM) and ordinary baryonic matter (OBM) distribution. Here, we use the MIT bag model equation of state to connect the density and pressure of SQM inside stars, whereas the linear equation of state $p_r =\alpha \rho-\beta$ connects the radial pressure and matter density caused by baryonic matter. The stellar model was developed under a background of $f(Q)$ gravity using the quadratic form of $f(Q)$. We utilized the Tolman-Kuchowicz ansatz [R. C. Tolman, Phys. Rev. 55 (1939) 364–373; B. Kuchowicz, Acta Phys. Pol. 33 (1968) 541] to find the solutions to the field equations under modified gravity. We have matched the interior solution to the external Schwarzschild spacetime in order to acquire the numerical values of the model parameters. We have selected the star Her X-1 to develop various profiles of the model parameters. Several significant physical characteristics have been examined analytically and graphically, including matter densities, tangential and radial pressures, energy conditions, anisotropy factor, redshirt, compactness, etc. The main finding is that there is no core singularity present in the formations of the star under investigation. The nature of mass and the bag constant $B_g$ have been studied in details through equi-mass and equi-$B_g$ contour. The maximum allowable mass and the corresponding radius have been obtained via $M-R$ plots.  

\end{abstract}

\maketitle

\section{Introduction}
The spatial structure of the universe's rapid expansion has drawn a lot of emphasis in the latest developments of cosmology and astronomical physics \cite{Pm,Rie}. Modern innovations in this cosmological period have shown novel ways to familiarise the essential and empirical changes for the fast evolution of the galaxy. Various findings could offer persuasive evidence of the rapid growth caused by extreme redshift supernova observations \cite{Rie2}, whereas massive formations \cite{Teg} and changes in the celestial microwave radiation \cite{Spe} present implicit support. An unidentified aspect known as dark energy $(DE)$, which sustains an intense adverse force, is responsible for the universe's accelerated expansion. Also, unexplained DE is believed to include around $68\%$ of the universe's overall energy. Therefore, it is necessary to make certain adjustments to the conventional theory in order to evaluate the occurrence of rapid growth. These sorts of trials encourage researchers to explore possibilities for modified or expanded theories of gravity that may be capable of illustrating scenarios when the general theory of relativity $(GR)$ generates unacceptable conclusions. Due to the constraints of $GR$, cosmologists are curious about analyzing modified gravitational theories. Some of these theories are $f(R),$ $f(G),$ $f(Q),$ $f(T),$ $f(R, G),$ $f(R, T),$  and $f(R, \phi)$ gravitational theories \cite{Zoy}-\cite{Zoy1}. The alterations of $GR$ seem enticing to explain the late-time of cosmic evolution and $DE$ difficulties. In addition, the various astronomical perspectives and concepts offered by these theories assist in elucidating the mysteries underlying the occurrence of the galaxy's rising expansion \cite{Cap2}. Scientists need to verify the reliability of these kinds of modified theories of gravity in all scales, like cosmological scales and astrophysical ones. It is reasonable to assume that altering the gravitational field action will have an impact on the astrophysical point of view. In the weak field limit, modified theories of gravity reduce to GR, whereas the strong field regimes may be able to distinguish between GR and its potential extensions. It is commonly known that relativistic compact objects (neutron stars) live in strong gravitational fields, so this kind of astrophysical object can be studied to check the possible deviation of the newly proposed modified gravity theory from Einstein's GR. Additionally, new phenomena that Einstein could not explain can be discovered in stellar astrophysics through this modified theory of gravity.\\
A hybrid star is an assumed particular kind of star in which a neutron star that is located at the center of a red giant or red supergiant, produced through an explosion of the massive with neutron star and extremely high density. Hybrid stars are yielded as an outcome of gravitational deformation when the nucleus of a star loses out of energy and is unable to sustain its weight despite the force of gravity. One of the universe's strangest and weirdest things is the hybrid object. They are typically connected to phenomena like eruptions of supernovae, cosmic rays, and bursts of gamma radiation. Researching dense stars may assist scientists in better understanding the properties of matter at very high densities and how energies and matter behave under very intense fields of gravity. According to the altered ideas, a hybrid stellar is an exclusive type of celestial object that occurs by the collapse of matter against the pressure of powerful gravitational forces, also defined by modified equations. One of the important characteristics of modified gravity is the ability to accommodate non-singular hybrid stars, which does not anticipate by the standard $GR$. The core of these non-singular giant stars is uniform and smooth and it is linked to the external geometry. According to research on the behavior of hybrid stars in modified gravity, the features of these structures can be quite distinct from those believed by $GR$. One of the important characteristics of modified gravity is the ability to accommodate non-singular hybrid stars, which does not anticipate by the standard $GR$. The core of these non-singular giant stars is uniform and smooth and it is linked to the external geometry. According to research on the behavior of compact stars in modified gravity, the features of these structures can be quite distinct from those believed by $GR$.\\
Plenty of researchers have implemented some important refinements to GR in the last couple of decades. In these beneficial amendments, one of the most intuitive and prominent theory is obtained by replacing the expression of Ricci scalar $R$ with an arbitrary function $f(R)$ \cite{Buch}. Such different models of gravity serve as essential for the accelerating proliferation of space give better explanation for the enigmatic composition of the cosmos. The fascinating theory that gained prominence in recent decades is symmetric teleparallel gravity \cite{Nes}, acknowledged as the $f(Q)$ theory. Jimenez et al. \cite{Jim} proposed the idea of $f(Q)$, in which the nonmetricity $Q$ essentially initiates the gravitational attraction. Studies into $f(Q)$ gravity are progressing efficiently, as have empirical obstacles to compare it to the conventional $GR$ interpretation. Lazkoz et al. \cite{Laz} established an intriguing collection of limitations on $f(Q)$ gravity by defining the $f(Q)$ Lagrangian as polynomial equations of the redshift $z$. According to these investigations, feasible $f(Q)$ models have coefficients similar to the $GR$ model namely  $\Lambda CDM$ model. They have checked the validity of these models at the background level to see if this new formalism offers any viable alternatives to explain the late-time acceleration of the universe. For this verification, they have used a variety of observational probes, such as the expansion rate data from early-type galaxies, Type Ia Supernovae, Quasars, Gamma Ray Bursts, Baryon Acoustic Oscillations data, and Cosmic Microwave Background distance priors. This innovative method offers an alternative viewpoint on developing a modified, observationally trustworthy gravity model. Apart from this, there is some work \cite{M,SH} based on the observational constraints in the background of $f(Q)$ gravity which gives the strong motivation to explore stellar models in this $f(Q)$ theory. Mandal et al. \cite{Man} investigated energy parameters for the power-law and nonlinear $f(Q)$ models that describe the visible behavior of the cosmos. Jimenez et al. \cite{J1} discussed the modified gravity theories built on nonlinear extensions of the nonmetricity scalar, and investigated several intriguing baseline cosmologies (such as accelerating solutions relevant to inflation and dark energy), and examined the response of cosmic disturbances. By giving the evolution equations and enforcing certain functional forms of the functions, such as power-law and exponential dependence of the nonminimal couplings, Harko et al. \cite{J8} investigated a number of cosmological applications. Mandal et al. \cite{Man2} reconstructed the appropriate structure of the $f(Q)$ function in $f(Q)$ gravity by employing cosmographic factors and also studied the different sorts of energy constraints for the exploration of logarithmic and polynomial functions in the $f(Q)$ gravity. Khyllep \cite{K1} explored the cosmic nature of power-law structure and the rapid evolution of matter perturbation in the modified $f(Q)$ gravity. Anagnostopoulos, et al. \cite{q1} proposed a novel model in the framework of $f(Q)$ gravity, which has the same number of free parameters to those of $\Lambda CDM$, however at a cosmological framework it gives rise to a scenario that does not have $\Lambda CDM$ as a limit. Frusciante \cite{q2} focused on a specific model in $f(Q)$ gravity which is indistinguishable from the $\Lambda$-cold-dark-matter model at the background level, while showing peculiar and measurable signatures at linear perturbation level. Lin and Zhai \cite{q3} explored the application of $f(Q)$ gravity to the spherically symmetric configurations and demonstrated the effects of$f(Q)$ by considering the external and internal solutions of compact stars. Ambrosio, et al., \cite{q4} constructed several perturbative corrections to the Schwarzschild solution for different choices of $f(Q)$, which in particular include a hair stemming from the now dynamical affine connection. De and Loo \cite{q5} proved that the energy conservation criterion is equivalent to the affine connection's field equation of $f(Q)$ theory.   \\
Astronomers have observed that the Tolman-Kuchowicz metric to be quite intriguing topic for studying the evolution of astronomical formations. Jasim et al. \cite{K2} investigated a singularity-free model for spherically symmetric anisotropic peculiar stars using the Tolman–Kuchowicz metric. In the setting of modified $f(R,G)$ gravity, Javed et al. \cite{K3} studied a variety of anisotropic star spheres and developed equations of motion that take into account anisotropic matter distribution and Tolman-Kuchowicz spacetime. Shamir and Naz \cite{K4} examined certain relativistic stellar object configurations for static spherically symmetric structures under modified gravity using the Tolman-Kuchowicz spacetime. Biswas et al. \cite{K6} offered a relativistic model of a static, spherically symmetric, anisotropic odd star based on Tolman-Kuchowicz metric potentials and they further employed the most basic version of the phenomenological MIT bag equation of state to characterize the distribution of SQM across the star system. Majid and Sharif \cite{K7} created an anisotropic model of strange stars in the context of massive Brans-Dicke gravity and used the MIT bag model to obtain the field equations for the Tolman-Kuchowicz ansatz. Within the context of Einstein-Gauss-Bonnet gravity in five dimensions, Bhar et al. \cite{K8} studied the distribution of anisotropic compact matter by solving the corresponding field equations using the inner geometry of Tolman-Kuchowicz spacetime.  Naz and Shamir \cite{K9} explored the effect of electric charge on static spherically symmetric star models in the presence of anisotropic matter distribution using the Tolman-Kuchowicz space-time and the simplified phenomenological MIT bag equation of state. Zubair et al. \cite{K10} introduced stellar models for anisotropic matter distribution under $f(T)$ gravity and generated matching conditions by combining the interior geometry of Tolman-Kuchowicz spacetime with exterior spacetimes. Saklany et al. \cite{K11} provided a simple description for modeling the coupling of dark energy with OBM by employing the super-dense pulsar PSRJ1614-2230 as the model star, and the field equations are solved in the stellar interior using the generalized framework of Tolman-Kuchowicz spacetime metric. The authors of the article \cite{pk1} examine the anisotropic stellar solutions admitting Finch-Skea symmetry (viable and nonsingular metric potentials) in the presence of some exotic matter fields. In the work \cite{pk2}, authors derived the exact solutions for the relativistic compact stars in the presence of two fields axion (Dante's Inferno model) and with/without the complex scalar field (with the quartic self-interaction) coupled to gravity. Recently, Astashenok, et al. \cite{Av1} investigated the Chandrasekhar mass limit of white dwarfs in various models of $f(R)$ gravity by taking two equations of state for stellar matter: the simple relativistic polytropic equation with polytropic index and the realistic Chandrasekhar equation of state. Astashenok along with his collaborators \cite{Av2} investigated the upper mass limit predictions of the baryonic mass for static neutron stars in the context of $f(R)$ gravity by using the most popular $R^2$ gravity model. Astashenok and Odintsov \cite{Av3} investigated realistic neutron stars in axion $R^2$ gravity and obtained the increase of star mass independent from central density for wide range of masses. The same authors \cite{Av4} investigated the equilibrium configurations of uniformly rotating neutron stars in $R^2$ gravity with axion scalar field for GM1 equation of state for nuclear matter. Some interesting work related to the stellar structures can be seen in \cite{Av5}-\cite{Av10}.  \\
Many researchers proposed the model of compact star in modified theory of gravity which has been discussed earlier. In this paper our goal is to obtain a hybrid star model in f(Q) gravity which can include the recent observation of different compact star. From our analysis, with the help of the mass radius profile we are able to attain the mass of different compact star in the f(Q) gravity which has been discussed in this paper and it is one of the most positive outcome of our present paper. To the best of our knowledge, this is first attempt to discuss the physical characteristics and maximum allowable mass of hybrid star in the background
of $f(Q)$ gravity. The arrangement of the current manuscripts is as follows: Section II deals with the basic formalism of $f(Q)$ theory of gravity. In Section III, we discuss the Tolman-Kuchowicz ansatz and MIT bag model equation of state. Matching condition has been investigated in Section IV. Section V deals with the mass, surface redshift and compactness factor. Mass radius relationship is presented in Section VI with details. The mass and bag constant by using colored plots are represented in Section VII. Sections VIII deals with the details discussion of physical analysis of considered stellar structures. Lastly, we conclude the outcome of our findings.

\section{Construction of $f(Q)$ gravity }\label{sec2}
Now, we introduce the action for $f(Q)$ gravity given by \cite{BeltranJimenez:2017tkd},
\begin{eqnarray}\label{x1}
S=\int\left[\frac{1}{2}f(Q)+\mathcal{L}_m\right]\sqrt{-g}d^{4}x,
\end{eqnarray}
where $f(Q)$ is a general function of $Q$, $g$ represents the determinant of the
metric $g_{\mu \nu}$ and $\mathcal{L}_m$ is the matter
Lagrangian density. The non-metricity tensor is given as,
\begin{eqnarray}
Q_{\alpha \mu \nu}=\nabla_{\alpha}g_{\mu \nu}=-L^{\rho}_{\alpha \mu}g_{\rho \nu}-L^{\rho}_{\alpha \nu}g_{\rho \mu},
\end{eqnarray}
where the following equations serve as representations for the non-metricity tensor's two independent traces:
 \begin{eqnarray}
 Q_{\alpha}=Q^{~\beta}_{\alpha ~\beta},~~ \tilde{Q}_{\alpha}=Q^{\beta}_{~~\alpha \beta},
 \end{eqnarray}and the deformation term is given by,
\begin{eqnarray}
L^{\alpha}_{\mu \nu}=\frac{1}{2}Q^{\alpha}_{\mu \nu}-Q^{~~~\alpha}_{(\mu \nu)},
\end{eqnarray}
whereas $Q $ is given as,
 \begin{eqnarray}
Q=-g^{\mu\nu}(L^{\alpha}_{\beta \nu}L^{\beta}_{\mu \alpha}-L^{\beta}_{\alpha \beta}L^{\alpha}_{\mu \nu})=-P^{\alpha \beta \gamma}Q_{\alpha \beta \gamma}.
 \end{eqnarray}
Here, $P^{\alpha \beta \gamma}$ is the non-metricity conjugate and the corresponding
tensor is written as
\begin{eqnarray}
P^{\alpha}_{~~\mu\nu}=\frac{1}{4}\left[-Q^{\alpha}_{~~\mu\nu}+2Q^{\alpha}_{(\mu\nu)}-Q^{\alpha}g_{\mu \nu}-\tilde{Q}^{\alpha}g_{\mu \nu}-\delta^{\alpha}_{(\mu}Q_{\nu})\right].
\end{eqnarray}
The field equation of $f(Q)$ gravity is obtained if we vary (\ref{x1}) with respect to $g_{\mu\nu}$ and it takes the following form:
\begin{eqnarray}
-\frac{2}{\sqrt{-g}}\nabla_a (\sqrt{-g}f_QP_{\mu \nu}^{\alpha})+f_Q(P_{\nu}^{\alpha\beta}Q_{\mu \alpha \beta}-2P^{\alpha \beta}_{~~\mu}Q_{\alpha\beta\nu})+\frac{1}{2}g_{\mu\nu}f=\kappa T_{\mu \nu}
\end{eqnarray}
where $f_Q=\frac{\partial f}{\partial Q}$ and
the energy-momentum tensor $T_{\mu \nu}$ is given by
\begin{eqnarray}\label{tmu1}
T_{\mu \nu}&=&-\frac{2}{\sqrt{-g}}\frac{\delta \sqrt{-g}\mathcal{L}_m}{\delta \sqrt{g_{\mu \nu}}},
\end{eqnarray}
Now, by altering the action in relation to the affine connection, the following equation can be obtained:
\begin{eqnarray}
\nabla_{\mu}\nabla_{\nu}(\sqrt{-g}f_QP^{\mu \nu}_{~~~\alpha})=0.
\end{eqnarray}
Within the framework of $f(Q)$ gravity, the field equations guarantee the conservation of the energy-momentum tensor, and given the choice of $f(Q)=Q$, the Einstein equations are retrieved.

\section{Modified Field Equation in $f(Q)$ gravity}

We have considered the following line element as:
\begin{equation}\label{line1}
ds^{2}-=e^{\nu}dt^{2}-e^{\lambda}dr^{2}-r^{2}(d\theta^{2}+\sin^{2}\theta d\phi^{2}),
\end{equation}
where, $\lambda$ and $\nu$ are functions of `r' and $0\leq r < \infty$. The metric co-efficients $\lambda$ and $\nu$, only depend on $r$. If both $\nu(r)$ and $\lambda(r)$ tend to $0$ as $r\rightarrow \infty$ , the spacetime will be asymptotically flat.\par
In the present article we have described a model of the hybrid star which is made up of normal baryonic matter having density $\rho$ along with the strange quark matter having density $\rho_q$ and for the sake of simplicity we have not considered the interaction between these two matters. For the presence of these two types of matter, the energy-momentum tensor is changed as follows:
\begin{eqnarray}
  T_0^0=\rho^{\text{eff}}=\rho+\rho_q ,\label{t1}\\
  T_1^1=-p_r^{\text{eff}}=-(p_r+p_q),\\
  \text{and}~~~
  T_2^2=T_3^3=-p_t^{\text{eff}}=-(p_t+p_q).\label{t3}
\end{eqnarray}
In the present scenario, $\rho$, $p_r$, and $p_t$ refer to the matter density, radial pressure, and transverse pressure generated by traditional baryonic matter, while $\rho_q$ and $p_q$ refer to the matter density and pressure developed by quark matter, respectively.\par
 Bhar \cite{Bhar:2015wma} also used the same technique to model a compact star in GR. Abbas and Nazar \cite{Abbas:2021uwt} recently used the same approach to model a hybrid star in minimally coupled $f(R)$ gravity. In our present article, our goal is to study the effect of the coupling parameter of $f(Q)$ gravity on the model of a hybrid star. \\
A crucial factor in the composition of ultra-dense strange quark particles is the incorporation of SQM in the fluid distribution. It has been hypothesized that the neutrons' phase change into bosons, hyperons, and SQM may occur at the core of the neutron star due to the immense pressure and density present there. According to Cameron's analysis \cite{cameron}, the hyperon must be produced inside the neutron star. Some nucleons may be converted into hyperons, which are more supportive energetically, as a result of extremely massive density and weak interaction. Quark matter, however, may also be present in the neutron star's interior. Due to the massive density and high central momentum conversion in the neutron star's core, the quarks become free of interaction. According to a review of the literature, the (u) and (d) quarks are currently undergoing strange matter transformations, and the entire quark matter also undergoes strange matter transformations \cite{Alcock:1986hz, Haensel:1986qb, Itoh:1970uw, Bodmer:1971we, Witten:1984rs}. As a result, the neutron star as a whole gets converted into a strange quark object \cite{Pagliara:2013tza}. Some other work related to the hybrid star can be found in \cite{Yan:2012mk, Schertler:1997za, Schertler:2000xq}.\\
We have the following field equations for a hybrid star in $f(Q)$ gravity using all the aforementioned expressions:
\begin{eqnarray}
\kappa(\rho+\rho_q)&=&\frac{e^{-\lambda}}{2r^2}\Big[2rf_{QQ}Q'(e^{\lambda} - 1) +
   f_Q\Big((e^{\lambda} - 1)(2 + r\nu') + (1 + e^{\lambda})r \lambda'\Big) +
   f r^2 e^{\lambda}\Big],   \label{fe1}\\
\kappa (p_r+p_q) &=&-\frac{e^{-\lambda}}{2r^2}\Big[2rf_{QQ}Q'(e^{\lambda} - 1) +
   f_Q\Big((e^{\lambda} - 1)(2 + r \lambda' + r\nu') - 2r\nu'\Big) +
   fr^2e^{\lambda}\Big], \label{fe2}\\
\kappa (p_t+p_q) &=& -\frac{e^{-\lambda}}{4r}\Big[-2rf_{QQ}Q'\nu' +
   f_Q\Big(2\nu'(e^{\lambda} - 2) - r\nu'^2 + \lambda'(2e^{\lambda} + r\nu') -
      2r\nu''\Big) + 2fre^{\lambda}\Big].\label{fe3}
\end{eqnarray}
where $\kappa=8\pi$ and $(^{\prime})$ represents the derivative with respect to the radial co-ordinate `$r$'. Now, let us choose a linear function for $f(Q)$ gravity, which is expressed as:
\begin{eqnarray}\label{g3}
f(Q)=mQ+n,
\end{eqnarray}
where `$m$' and `$n$' are characteristics without dimensions. 
The expression of $Q$ is described by \cite{Lin:2021uqa},
\begin{eqnarray}\label{g2}
Q=\frac{1}{r}(\nu'+\lambda')(e^{-\lambda}-1).
\end{eqnarray}

\section{Model of Hybrid Star in $f(Q)$ gravity}\label{sec3}
To obtain the model of the hybrid star, let us use the well-known Tolman-Kuchowicz {\em ansatz} \cite{Tolman1939, K68} given by,
\begin{eqnarray}\label{eq13}
\nu(r) &=& Br^2+2\ln D,\\ \label{eq14}
\lambda(r)&=& \ln(1 + ar^2 + br^4),
\end{eqnarray}
where $D$ is a free of dimensions parameter and $a$, $B$, and $b$ are parameter values that are constant having units of km$^{-2}$, km$^{-2}$, and km$^{-4}$, respectively. The metric potentials chosen in this paper are well-motivated since they provide a model which does not suffer from any kind of singularity.\\
To close the system we have to choose one extra constraint, i.e., a well-motivated relation between the radial pressure $p_r$ and density $\rho$ of normal baryonic matter is needed. There are several choices to describe a relation between $p_r$ and $\rho$. For our present model, we have chosen a linear equation of state given by 
\begin{eqnarray}\label{eos1}
 p_r &=& \alpha \rho-\beta,
 \end{eqnarray}
where $0<\alpha<1$ with $\alpha\neq 1/3.$ and $0<\beta$. Many authors have used this EoS to model the compact star which can be found in Refs. \cite{Sharma:2007hc, Ngubelanga:2015dih, Sarkar:2020tbv, Abbas:2021uwt}. Our work is well motivated by these articles.\\
Let's further assume that the MIT bag model equation of state provides the pressure-matter density relation for quark matter as follows: \cite{Cheng:1998na, Witten:1984rs},
\begin{eqnarray}\label{eos2}
  p_q &=& \frac{1}{3}(\rho_q-4B_g),
\end{eqnarray}
where $B_g$ is the bag constant of units MeV/fm$^3$ \cite{Chodos:1974je}.
Now solving the equations (\ref{fe1})-(\ref{fe3}) with the help of (\ref{g3})-(\ref{eos2}), we obtain:

\begin{eqnarray}
    \rho &=& 
\frac{1}{4 \pi  (3 \alpha -1) (a r^2+b r^4+1)^2} \Big[a^2 (r^4 (12 \pi  \beta +16 \pi  B_g-n)-2 m r^2)+a (m (-4 b r^4+3 B r^2-3) +3 B (b m r^4+m) \nonumber\\&&  -2 r^2 (b r^4+1) (n-4 \pi  (3 \beta +4 B_g)))-b r^2 (b r^4+2) (r^2 (n-16 \pi  B_g)+2 m)+12 \pi  \beta  (b r^4+1)^2+ 16 \pi  B_g -n\Big],
\end{eqnarray}

\begin{eqnarray}
    p_r &=& \frac{1}{4 \pi  (3 \alpha -1) (a r^2+b r^4+1)^2} \Big[a^2 (r^4 (4 \pi  (\beta +4 \alpha  B_g)-\alpha  n)-2 \alpha  m r^2)+ \nonumber\\&& \alpha  a (m (-4 b r^4+3 B r^2-3)-2 r^2 (b r^4+1)  (n-16 \pi  B_g))+8 \pi  a \beta  r^2 (b r^4+1)+ \nonumber\\&& \alpha  (3 B (b m r^4+m)-b r^2 (b r^4+2) (r^2 (n-16 \pi  B_g)+2 m)+16 \pi  B_g-n+4 \pi  \beta  (b r^4+1)^2\Big],
\end{eqnarray}

\begin{eqnarray}
   p_t &=& \frac{1}{8 \pi  (3 \alpha -1) (a r^2+b r^4+1)^2} \Big[-a^2 r^2 (2 r^2 (\alpha  n-4 \pi  (\beta +4 \alpha  B_g))+(\alpha +1) m)+a (-2 \alpha  (2 r^2 (b r^4+1) (n-16 \pi  B_g) \nonumber\\&& +m (b r^4+3))-2 b m r^4+16 \pi  \beta  r^2 (b r^4+1)+(3 \alpha -1) B^2 m r^4+(3 \alpha +1) B m r^2) \nonumber\\&& -b^2 r^6 (2 r^2 (\alpha  n-4 \pi  (\beta +4 \alpha  B_g)) +(\alpha +1) m)+b r^2 (B m r^2 ((3 \alpha -1) B r^2+2) \nonumber\\&& +4 r^2 (4 \pi  (\beta +4 \alpha  B_g)-\alpha  n)-11 \alpha  m+m)+8 \pi  \beta +3 \alpha  B^2 m r^2-B^2 m r^2+6 \alpha  B m+32 \pi  \alpha B_g-2 \alpha  n\Big],
\end{eqnarray}
and the anisotropic factor $\Delta$ can be gained as,
\begin{eqnarray}
   \Delta=p_t-p_r=\frac{m r^2 (a^2+a r^2 (2 b+B^2)-a B+b (r^4 (b+B^2)-2 B r^2-1)+B^2)}{8 \pi  (a r^2+b r^4+1)^2}.
\end{eqnarray}
Consequently, the components related to the SQM are as follows:
\begin{multline}
    \rho_q = \frac{1}{16 \pi  (3 \alpha -1) (a r^2+b r^4+1)^2}\Big[a^2 r^2 (r^2 (3 (\alpha +1) n-16 \pi  (3 \beta +4 B_g))+6 (\alpha +1) m)+2 a (m (9 \alpha +6 (\alpha +1) b r^4\\-6 B r^2+3) +r^2 (b r^4+1) (3 (\alpha +1) n-16 \pi  (3 \beta +4 B_g)))-64 \pi  b^2 B_g r^8+6 \alpha  b^2 m r^6+6 b^2 m r^6+3 \alpha  b^2 n r^8+3 b^2 n r^8\\-12 B (b m r^4+m)-128 \pi  b B_g r^4+30 \alpha  b m r^2+6 b m r^2+6 \alpha  b n r^4+6 b n r^4-48 \pi  \beta  (b r^4+1)^2-64 \pi  B_g+3 \alpha  n+3 n\Big],
\end{multline}
\begin{multline}
   p_q = \frac{1}{16 \pi  (3 \alpha -1) (a r^2+b r^4+1)^2}\Big[(a^2 r^2 (r^2 ((\alpha +1) n-16 \pi  (\beta +4 \alpha  B_g))+2 (\alpha +1) m)+2 a (m (3 \alpha +2 (\alpha +1) b r^4\\ -2 B r^2+1)+r^2 (b r^4+1) ((\alpha +1) n-16 \pi  (\beta +4 \alpha  B_g)))-64 \pi  \alpha  b^2 B_g r^8+2 \alpha  b^2 m r^6+2 b^2 m r^6+\alpha  b^2 n r^8+b^2 n r^8-4 B (b m r^4+m)\\-128 \pi  \alpha  b B_g r^4+10 \alpha  b m r^2+2 b m r^2+2 \alpha  b n r^4+2 b n r^4-16 \pi  \beta  (b r^4+1)^2-64 \pi  \alpha  B_g+\alpha  n+n)\Big].
\end{multline}

Our next objective is to use various physical acceptance tests to examine the current model's reliability. Those will be discussed in the coming sections.

\section{Exterior Spacetime and boundary conditions}
The material content that threads the star's interior must be confined between the centre and the boundary. The so-called junction conditions at the surface of the structure must be examined in order to ensure the restriction of this matter distribution. This process is carried out in GR by using the well-known Israel-Darmois \cite{is1,is2} matching requirements. The vacuum Schwarzschild solution \cite{sc1} is used to characterize external spacetime in this case as we are working with the uncharged fluid sphere and it is given by the following line element:
\begin{eqnarray}
ds^{2}&=&(1-\frac{2M}{r})dt^{2}-(1-\frac{2M}{r})^{-1}dr^{2}-r^{2}(d\theta^{2}+\sin^{2}\theta d\phi^{2}),
\end{eqnarray}
where `$M$' denotes the total mass within the boundary of the compact star. The continuations of the first and second fundamental forms at the boundary give the following relations:
\begin{eqnarray}
1-\frac{2M}{R}&=&e^{BR^{2}+2\ln D},\label{u1}\\
(1-\frac{2M}{R})^{-1}&=&1 + aR^2 + bR^4,\\
\frac{M}{R^{2}}&=&BRe^{BR^{2}+2\ln D},
\end{eqnarray}
and \begin{eqnarray}\label{u3}
    p_r(r=R)=0.
\end{eqnarray}
Resolving the aforementioned mathematical equations (\ref{u1})-(\ref{u3}), we get the following relations:
\begin{eqnarray}
    B &=& \frac{M}{R^3}(1 - 2\frac{M}{R})^{-1},\\
    D &=& e^{-B R^2/2}\sqrt{(1 - 2\frac{M}{R})},\\
    a &=& \frac{1}{R^2}((1 - 2\frac{M}{R})^{-1} - 1 - bR^4),\\
    \beta &=& 
 \frac{1}{4 \pi (1 + a R^2 + b R^4)^2} \Big(n - 16 Bg \pi + 
     b R^2 (2 m + (n - 16 Bg \pi) R^2) (2 + b R^4) - 
     3 B (m + b m R^4) \nonumber\\&& + a^2 (2 m R^2 + (n - 16 Bg \pi) R^4) + 
     a (2 (n - 16 Bg \pi) R^2 (1 + b R^4) + 
        m (3 - 3 B R^2 + 4 b R^4))\Big) \alpha
\end{eqnarray}

\begin{table*}[t]
\centering
\caption{The corresponding numerical values of $a,\,B$ and $D$ for some discriminate stellar spheres by undertaking $b=0.04\times 10^{-5}$~km$^{-4}$.}\label{table1}
\begin{tabular}{@{}ccccccccccccc@{}}
\hline
Star & Observed mass & Observed radius & Estimated  & Estimated &  $a$&$B$&$D$\\
& $M_{\odot}$ & km. & mass ($M_{\odot}$) & radius (km.)& $km^{-2}$ & $km^{-2}$\\
\hline
Her X-1 \cite{Abubekerov:2008inw}& $0.85 \pm 0.15$ & $8.1 \pm 0.41$ & 0.85 & 8.5 & 0.00576265 & 0.00289578& 0.756246       \\
EXO 1785-248 \cite{Ozel:2008kb} & $1.3 \pm 0.2$ & $8.849 \pm 0.4 $ &1.4 &8.85& 0.0111404 &0.00558588&0.586810 \\
Vela X-1 \cite{Rawls:2011jw} & $1.77 \pm 0.08$ & $9.56 \pm 0.08$    & 1.77 & 9.5& 0.0134864 & 0.00676124 & 0.494630     \\
PSR J1614-2230 \cite{Demorest:2010bx} & $1.97 \pm 0.04$ & $9.69 \pm 0.2$   & 1.97& 9.7& 0.0158465 & 0.00794205 &0.435751 \\
LMC X-4 \cite{Rawls:2011jw} & $1.04 \pm 0.09$ & $8.301 \pm 0.2$ & 1.04 & 8.3 & 0.00848444& 0.00425600 & 0.685691  \\
SMC X-4 \cite{Rawls:2011jw} & $1.29 \pm 0.05$ & $8.831 \pm 0.09$ & 1.29 & 8.8 & 0.0098081 & 0.00491954 & 0.622699 \\
PSR J1903+327 \cite{Freire:2010tf} & $1.667 \pm 0.021$ & $9.438 \pm 0.03$ & 1.67 & 9.4 &0.012428&0.00623168&0.523832  \\
4U 1538-52 \cite{Rawls:2011jw}& $0.87 \pm 0.07$ & $7.866 \pm 0.21$ & 0.87 & 7.8 & 0.00803612 &0.00403023&0.724610  \\
4U 1820-30 \cite{Guver:2008gc} & $1.58 \pm 0.06$&  $9.316 \pm 0.086$  & 1.58 & 9.3&0.0115823&0.00580843&0.549392  \\
Cen X-3 \cite{Rawls:2011jw} & $1.49 \pm 0.08$ & $9.178 \pm 0.13$ & 1.49 & 9.2& 0.0107751& 0.00540449& 0.574909 \\
\hline
\end{tabular}
\end{table*}

\section{Mass, Surface redshift and Compactness}
\begin{figure}[htbp]
    \includegraphics[scale=.47]{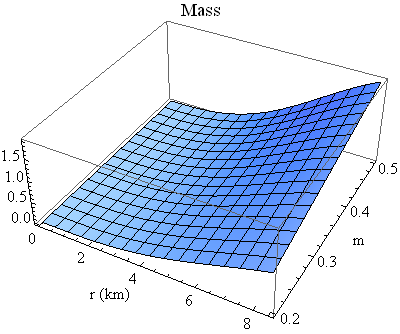}
        \includegraphics[scale=.47]{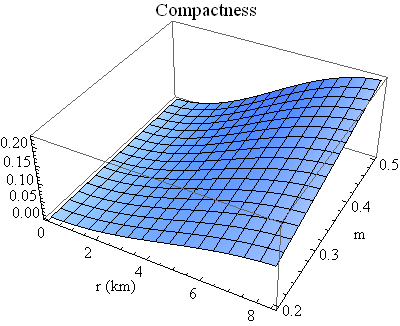}
        \includegraphics[scale=.47]{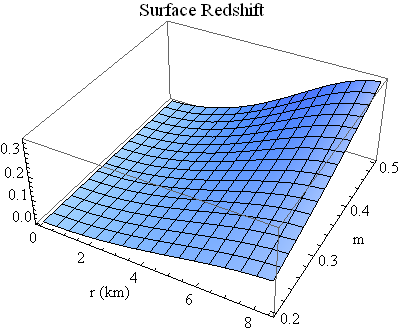}
       \caption{The graphical analysis of $\mathbf{m}(r)$, $z_s$, and $u(r)$ against `r' \label{mass1a}}
\end{figure}

The mass function $\mathbf{m}(r)$ is defined as 
\begin{eqnarray}
    \mathbf{m}(r) &=&\int _0^r 4\pi \rho(x)x^2 dx,\nonumber\\,
&=&\frac{1}{6 (3 \alpha -1)}\Big[\frac{9 \sqrt{2} m (B (\sqrt{a^2-4 b}-a)+b) \tan ^{-1}(\frac{\sqrt{2} \sqrt{b} r}{\sqrt{a-\sqrt{a^2-4 b}}})}{\sqrt{b} \sqrt{a-\sqrt{a^2-4 b}} \sqrt{a^2-4 b}}+\frac{3 m r}{a r^2+b r^4+1}\nonumber\\&&-\frac{9 \sqrt{2} m (b-B (\sqrt{a^2-4 b}+a)) \tan ^{-1}(\frac{\sqrt{2} \sqrt{b} r}{\sqrt{\sqrt{a^2-4 b}+a}})}{\sqrt{b} \sqrt{\sqrt{a^2-4 b}+a} \sqrt{a^2-4 b}}-2 r^3 (n-4 \pi  (3 \beta +4 B_g))-12 m r\Big]
\end{eqnarray}
Fig.~\ref{mass1a} displays the mass function profile. It is evident from the figure that there are no singularities in the mass function, which increases monotonically having the value zero at its centre.\\
The surface redshift $z_s$ is a crucial observable parameter that links the mass and the radius of a compact star and it is defined by the following formula:
\begin{eqnarray}
 z_s=(1-2\mathbf{m}(r)/r)^{-1/2}-1
 \end{eqnarray}
The surface redshift $z_s$ in Fig.~\ref{mass1a} exhibits a monotonic increasing behavior towards the boundary, reaching its maximum value at the boundary of the object. The values stated for $z_s$ in this paper are below the maximum values, despite the fact that Ivanov's research \cite{Ivanov:2002xf} shows that the value of $z_s$ in the presence of anisotropic fluids exceeds the Buchdahl constraint \cite{Buchdahl:1959zz}.
For our current model, the compactness factor is calculated as $u(r) = \mathbf{m}(r)/r$. To categorize compact objects as (i) regular stars ($u\sim 10^{-5}$), (ii) white dwarfs  ($u\sim 10^{-3}$), (iii) neutron stars  ($0.1<u<0.25$), (iv) ultra-compact star ($0.25<u<0.5$), and (v) black holes ($u=0.5$), the compactness factor is crucial. Fig.~\ref{mass1a} depicts the compactness profile for our current model, which is a monotonically increasing function of `r'.

\section{Mass Radius Relationship}
\begin{figure}[htbp]
    \includegraphics[scale=.5]{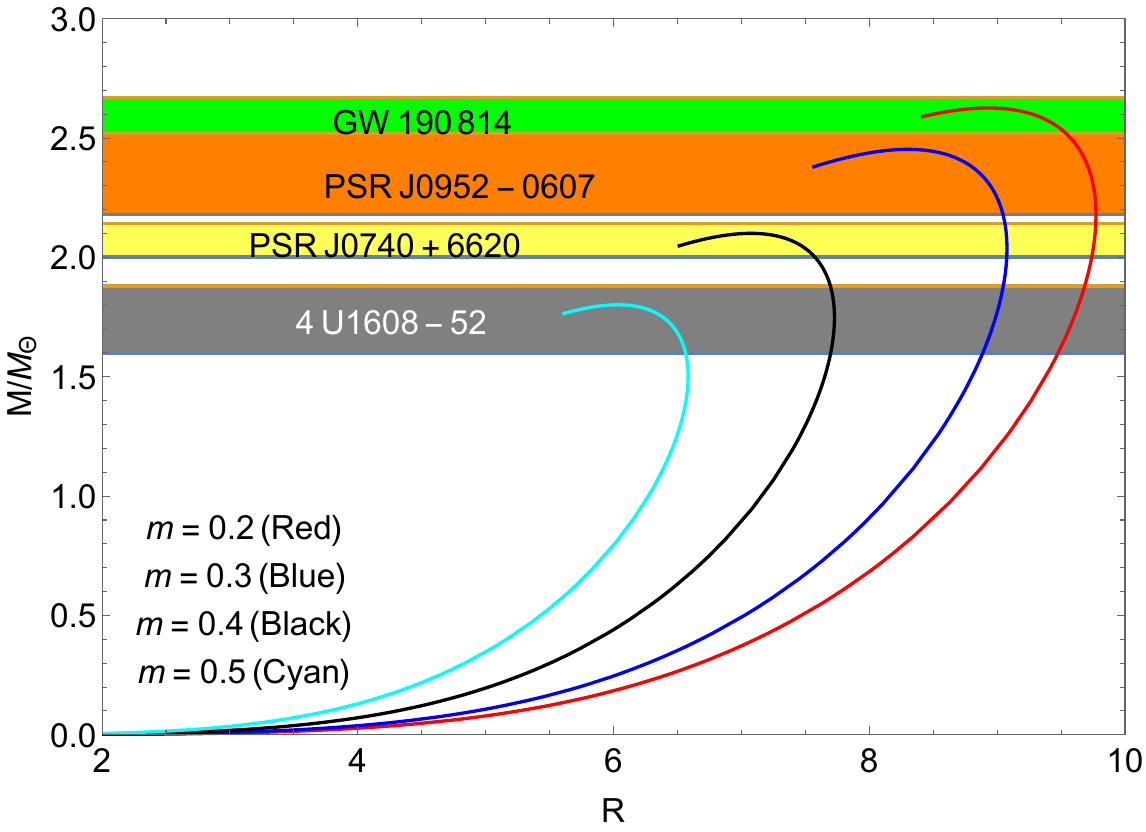}
       \caption{Mass-Radius relationship are shown \label{mass1b}}
\end{figure}

In this section, we are interested to find the maximum allowable mass for different values of $m$. As $m$
increases, the predicted masses cover a wider range of observed values which can be shown in fig.~\ref{mass1b}. An increase in $m$ is accompanied by a decrease in mass and radii, which is clear from the figure. From literature, we have chosen four different compact stars GW 190814 with mass $2.50–2.67~M_{\odot}$, PSR J0952-0607 with mass  $(2.35\pm 0.17)M_{\odot}$, PSR J0740+6620 with mass $(2.08\pm 0.07)M_{\odot}$ and 4U 1608-52 with mass $(1.74 \pm 0.14)M_{\odot}$. It is possible to generate stellar structures with masses closer to the above compact star for different values of $m$ which has been presented in Table~\ref{tb3a}. 

\begin{table*}[t]
\centering
\caption{ Maximum mass and the corresponding radius for different values of $m$ }\label{tb3a}
\begin{tabular}{@{}ccccccccccccc@{}}
\hline
$m$& Maximum mass $M(M_{\odot})$ && Corresponding radius (in km.) && Matched with the mass of the compact star \\
\hline
0.2& 2.62 &&    9.18 &&  GW 190814 \cite{LIGOScientific:2020zkf}\\
0.3& 2.4&&   8.6 && PSR J0952-0607 \cite{Romani:2022jhd} \\
0.4& 2.09 &&  7.3 && PSR J0740+6620 \cite{Fonseca:2021wxt}\\
0.5 & 1.8 &&    6.2 && 4U 1608-52 \cite{Guver:2010td} \\
\hline
\end{tabular}
\end{table*}

\section{Measurements of Mass and Bag Constant with the help of contour plots}

\begin{figure}[htbp]
    \includegraphics[scale=0.4]{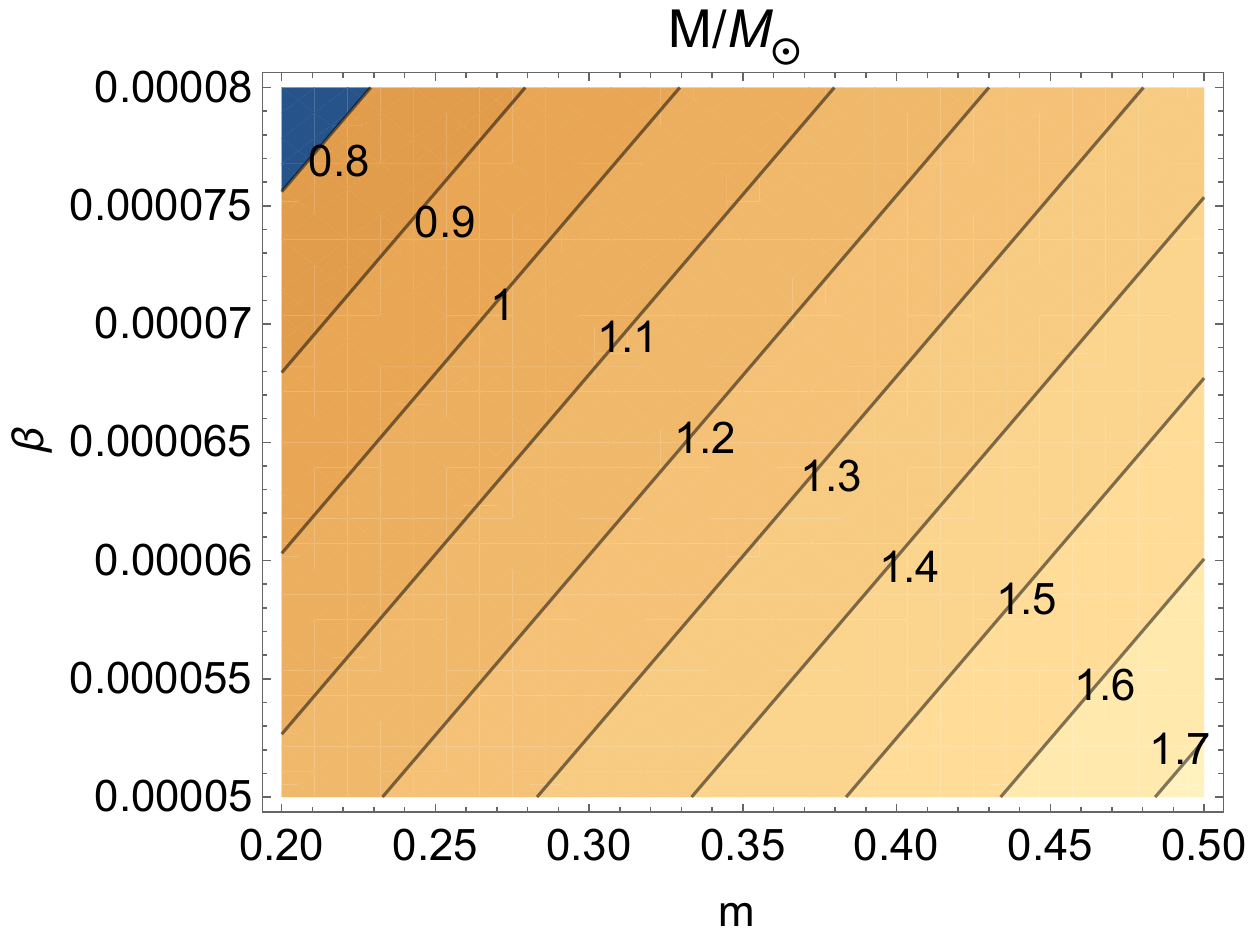}
       \caption{Equi-mass in $m-\beta$ plane \label{mass1c}}
\end{figure}

\begin{figure}[htbp]
    \includegraphics[scale=.4]{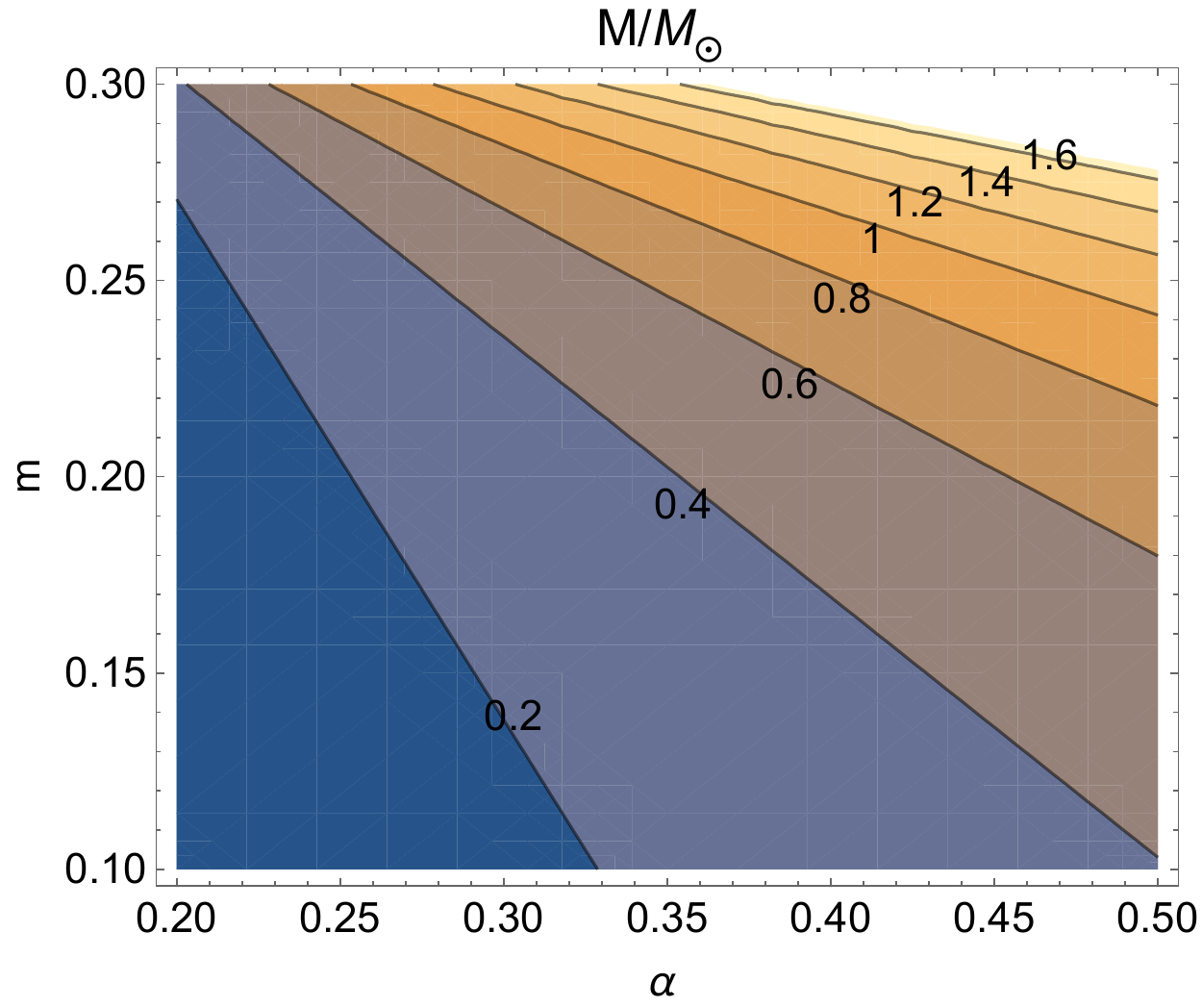}
      \includegraphics[scale=.4]{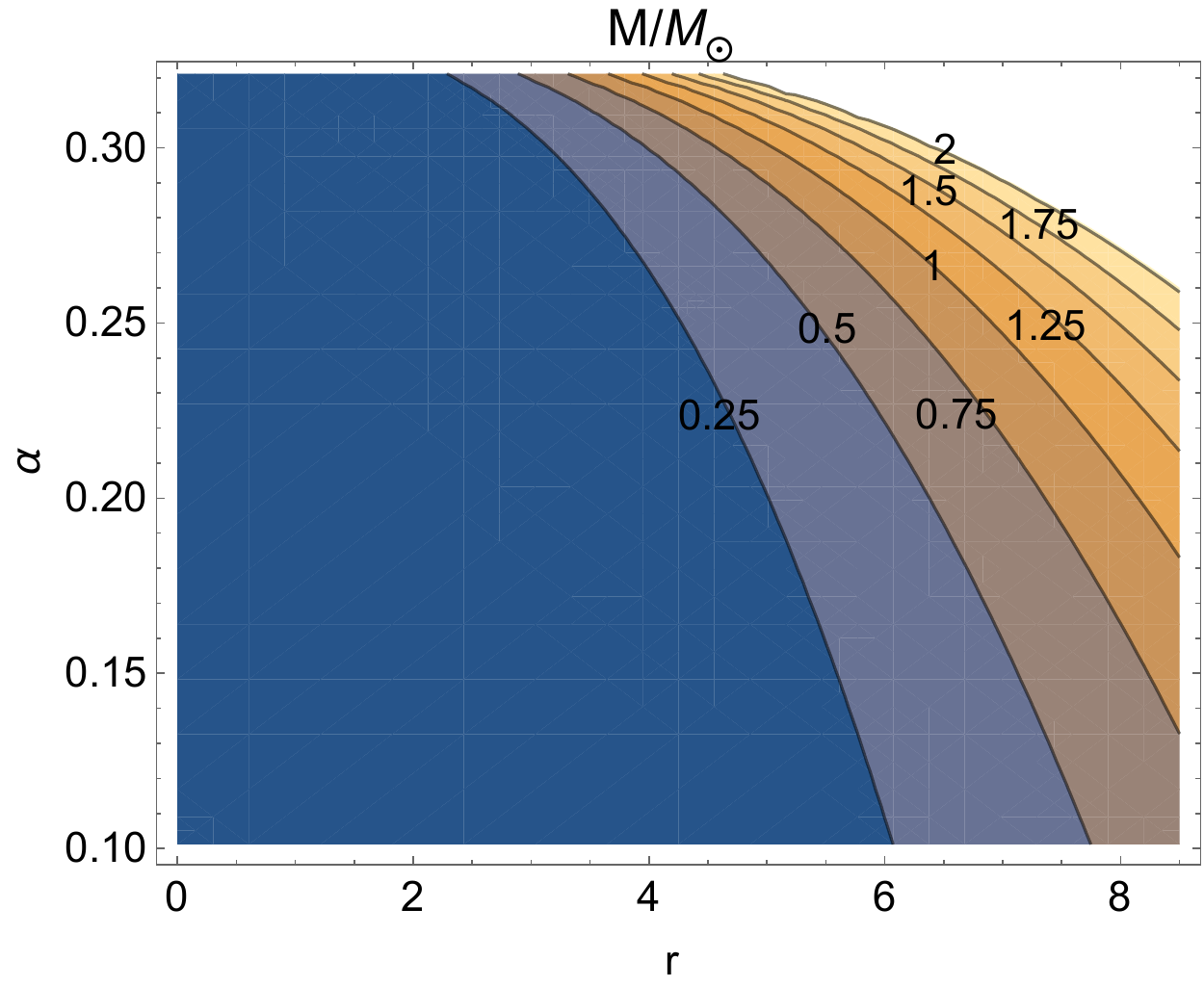}
       \caption{(Left) equi-mass in $\alpha-m$ and (right) equi-mass in $r-\alpha$ \label{mass1d}}
\end{figure}

\begin{figure}[htbp]
    \includegraphics[scale=0.4]{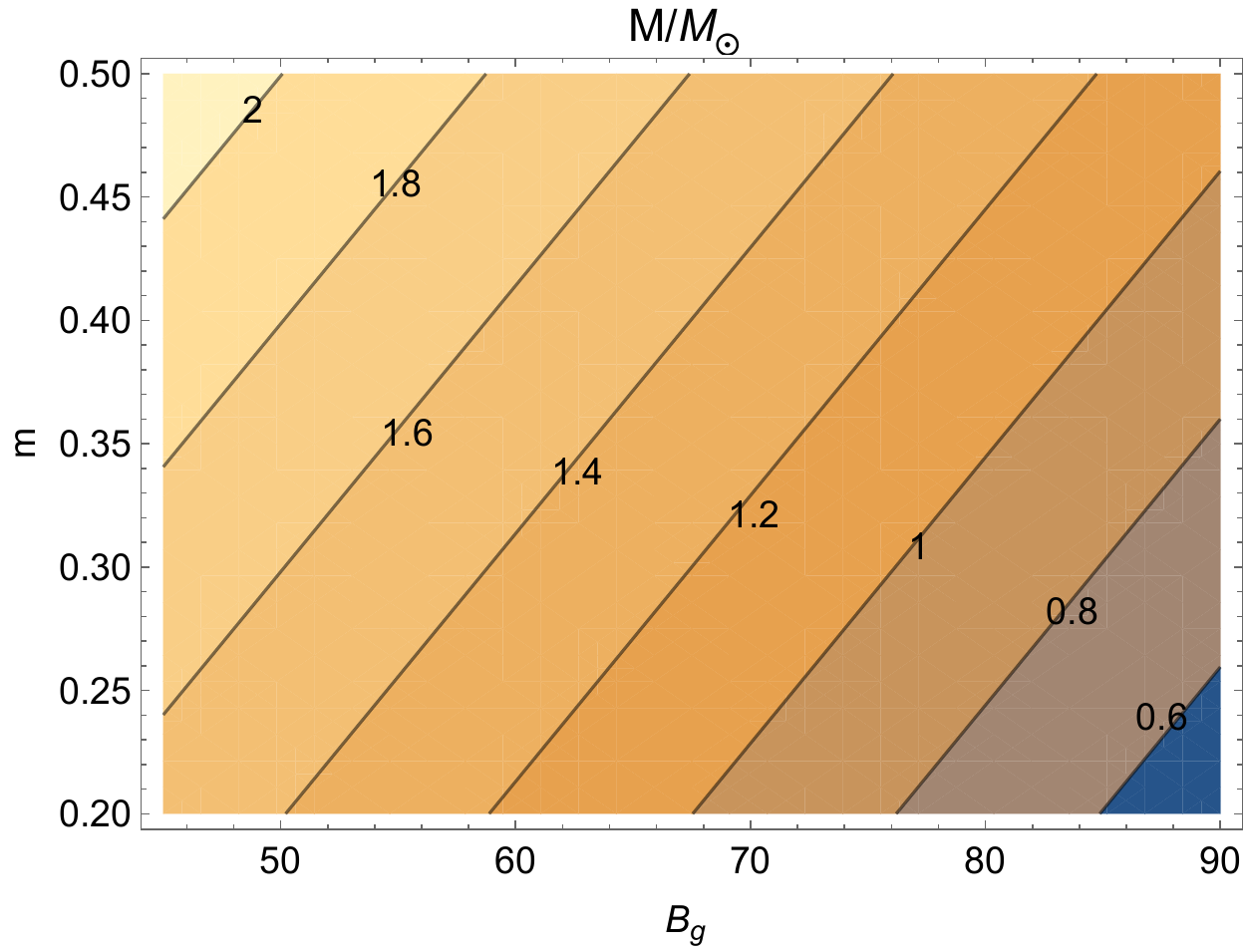}
      \includegraphics[scale=0.4]{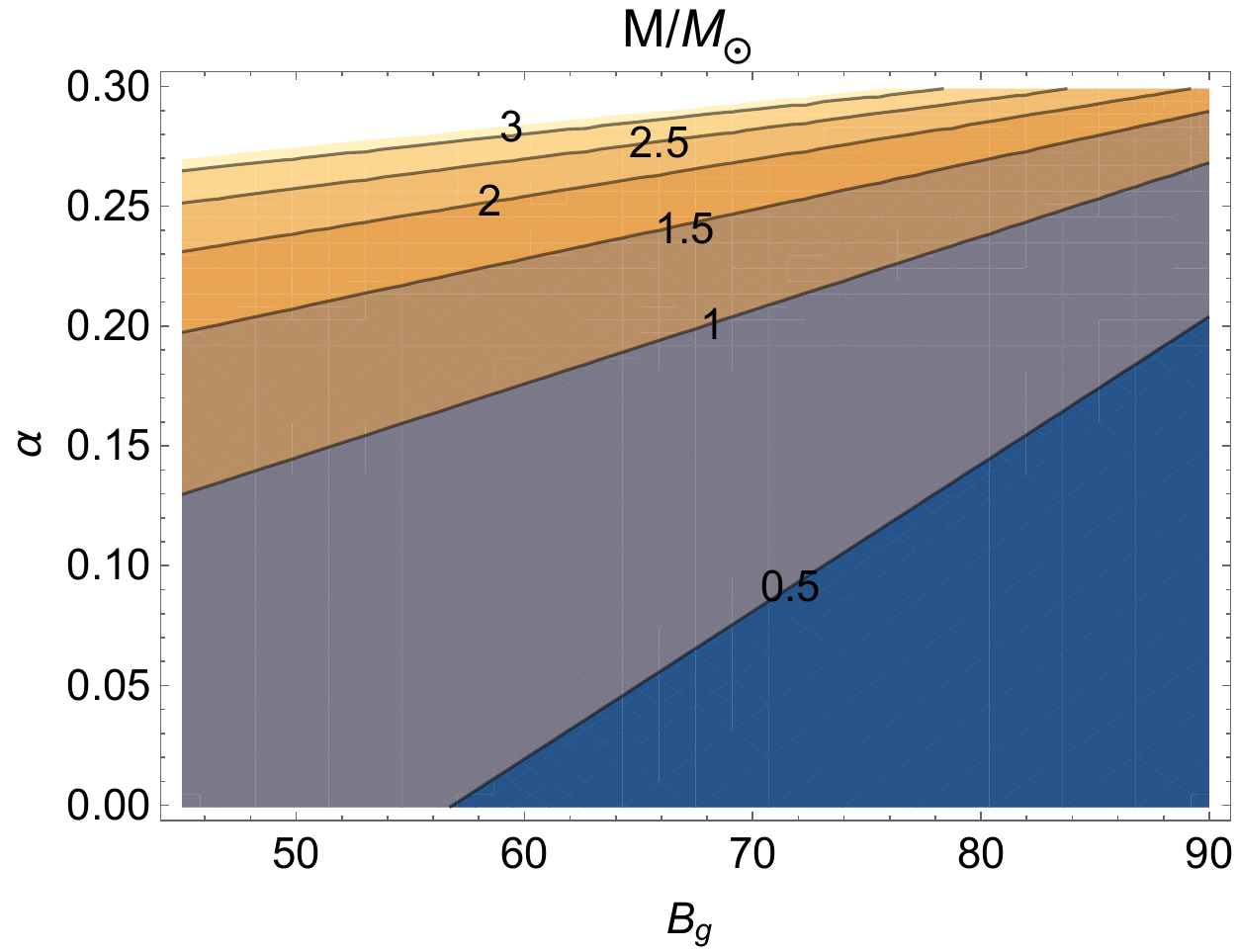}
       \caption{(Left) equi-mass in $B_g-m$ and (right) equi-mass in $B_g-\alpha$\label{mass1e}}
\end{figure}

\begin{figure}[htbp]
      \includegraphics[scale=.4]{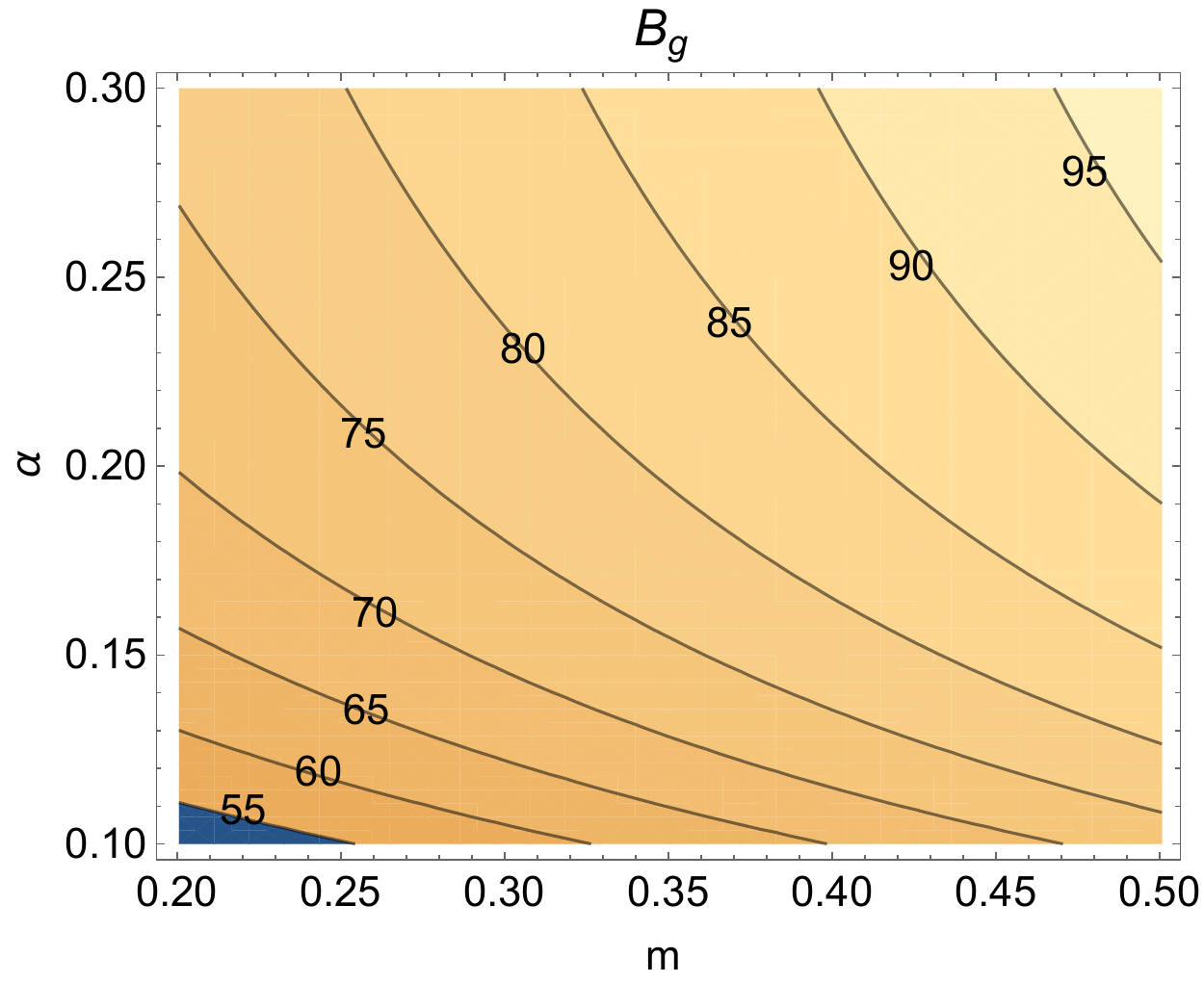}
      \includegraphics[scale=.4]{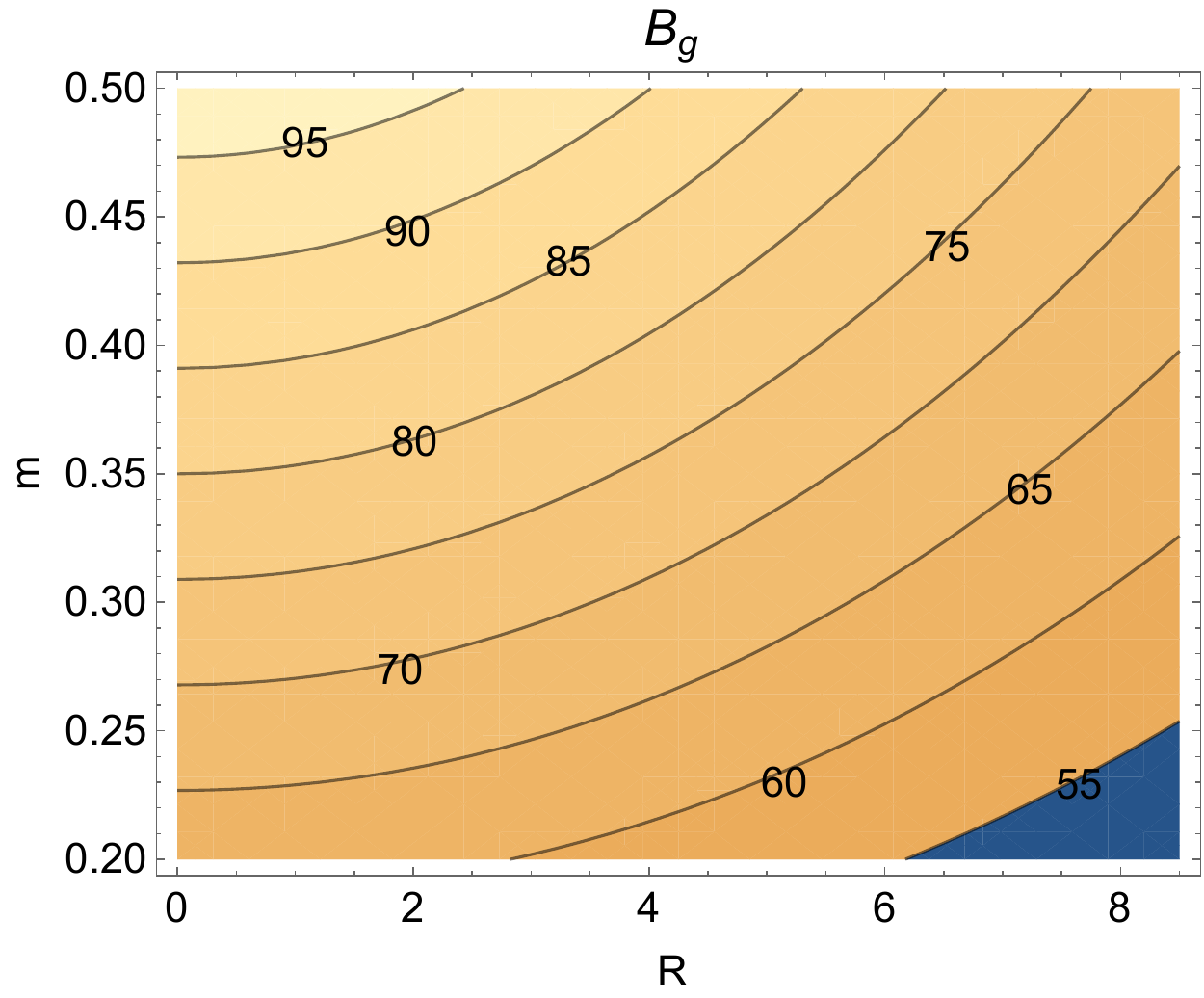}
       \caption{(Left) equi-$B_g$ in $m-\alpha$ and (right) equi-$B_g$ in $R-m$\label{mass1f}}
\end{figure}

From Fig.~\ref{mass1c} to Fig.~\ref{mass1f}, we analyzed the variation of mass and the bag constant with the help of contour plots.\par
  \begin{itemize}
      \item The equi-mass contours are shown in  the $m-\beta$ plane in Fig.~\ref{mass1c} by keeping $\alpha,\,n,\,r$ and $B_g$ fixed. The figure indicates that for a fixed value of $\beta$, the value of mass increases for an increasing value of $m$. In contrast, with a constant $m$, the value of mass falls as $\beta$ grows. 

\item The equi-mass contours are displayed in the $\alpha-m$ plane in the left panel of Fig.~\ref{mass1d} by retaining the variables $\beta,\,n,\,r$ and $B_g$ fixed. According to the picture, with a constant value of $\alpha$, the value of mass rises as $m$ increases. With a fixed amount of $m$, however, the value of mass grows as $\alpha$ increases.\par
In the right panel of Fig.~\ref{mass1d}, we have drawn the equi-mass contours in $r-\alpha$ plane taking $\beta,\,m,\,n$ and $B_g$ fixed. It can be seen that for a fixed value of $r$, the value of mass rises as $\alpha$ increases. Also, for a fixed value of  $\alpha$, the value of mass increases as $r$ increases.

\item In the left panel of Fig.~\ref{mass1e}, the equi-mass contours are displayed in the $B_g-m$ plane by keeping the variables $\beta,\,n,\,r$ and $\alpha$ fixed. According to the figure, with a constant value of $B_g$, the value of mass grows as $m$ increases. However, with a given quantity of $m$, the value of mass decreases as $B_g$ increases. We can see that, the mass takes a higher value for the lower value of  the bag constant $B_g$. \par
In the right panel of Fig.~\ref{mass1e}, the equi-mass contours are shown in the $B_g-\alpha$ plane by keeping the variables $\beta,\,n,\,r$ and $m$ fixed. One can see that, with a constant value of $B_g$, the value of mass grows as $\alpha$ increases. However, given a constant amount of $\alpha$, the value of mass falls as $B_g$ grows. 

\item The left panel of Fig.~\ref{mass1f} we show the equi-$B_g$ contours in the $m-\alpha$ plane by keeping the variables $\beta,\,n,\,r$ and $m$ fixed. This figure implies that with a constant value of $m$, the value of the bag constant increases as $\alpha$ increases. Similarly, for a fixed value of  $\alpha$,  the value of $B_g$ increases as $m$ grows. On the other hand, the right panel of Fig.~\ref{mass1f} shows the equi-$B_g$ contour in the $R-m$ plane. Keeping $R$ fixed, the value of bag constant $B_g$ increases as $m$ grows, and by keeping $m$ fixed, the value of $B_g$ decreases as $R$ increases. Interestingly, one can note that for our chosen range of $m$ and $\alpha$ in the left figure and for a chosen range of $R$ and $m$ in the right figure we have achieved very interesting and physically reasonable values for the bag constant $B_g$ which is very much consistent with the CERN data about quark-gluon plasma (QGP) as well as compatible with the RHIC preliminary results \cite{h1,h2}. Witten's conjecture successfully explains the non-interacting, mass-less quarks with $B_g$ values between $57$ and $94$ $MeV/fm^3$, which has already been demonstrated by Farhi and Jaffe \cite{farhi}.
\end{itemize}

\section{Physical Analysis}
We have discussed the analysis of the hybrid star model for a specific range of $m$ by fixing $n$ in this section. To check the behavior of the physical parameters and ensure the viability of the solution, we have chosen $m$ lies between $10$ to $15$ for our current article. The acquired solutions for the hybrid star model need to be put to the test under a number of different physical conditions, each of which will be addressed separately in this section. To create all of the curves of different model parameters, we utilized the stellar structures whose mass and radius are shown in Table.~\ref{table1}.
\begin{figure}[htbp]
        \includegraphics[scale=.3]{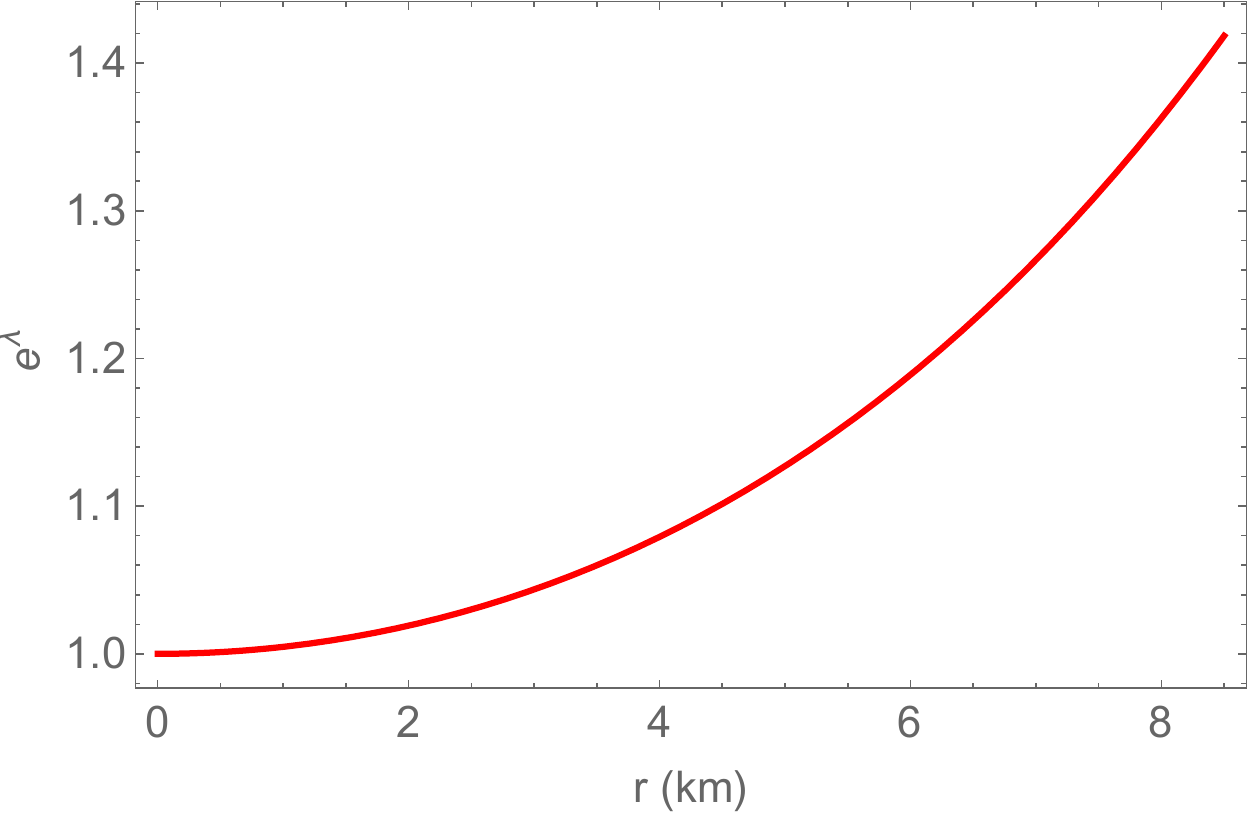}
          \includegraphics[scale=.3]{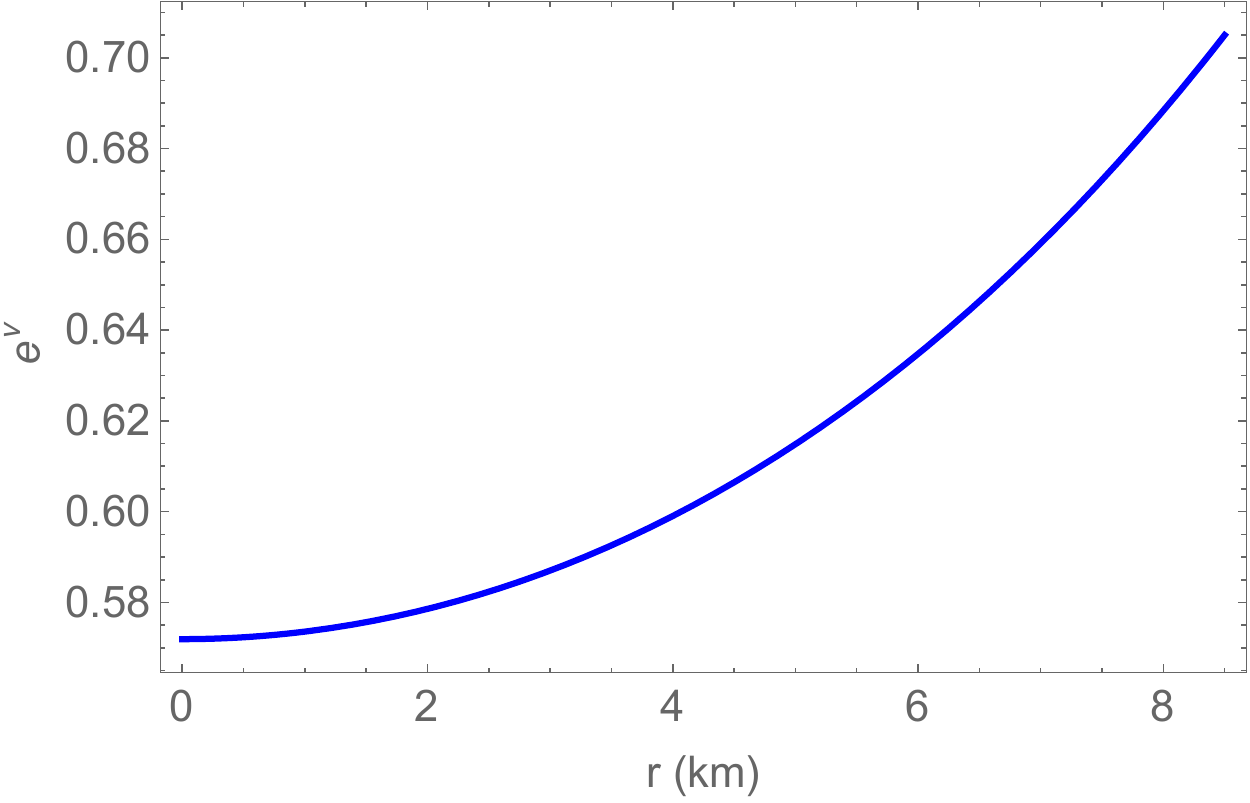}
       \caption{Metric coefficients\label{metric}}
\end{figure}
\subsection{Metric Potentials}
Both metric potentials are singularity-free within the boundary of the star. Additionally, $e^{\nu(0)}=D^2$, a non-zero constant, and $e^{-\lambda(0)}=1$ for our current stellar model. The derivative of the metric coefficients results in the expressions $(e^{\lambda})'=2ar+4br^3$, $(e^{\nu})'=2BD^2re^{Br^2}$. At the core of the star, the derivative of the metric potentials equals zero. Additionally, they are continuous and monotonic increasing inside the star as shown in Fig.~\ref{metric}. At the boundary, the metric components of the external Schwarzschild line element are perfectly aligned to the interior metric potentials, which will be addressed later.
\subsection{Nature of pressure, density and anisotropic factor}
The behavior of the three most important significant features of the model —matter density, radial pressure, and tangential pressure—is examined and analyzed in this subsection. We additionally examine the function that the anisotropy factor Delta plays inside the stellar sphere. It is well established that any compact object describing the interiors of stars should not have any physical or mathematical singularities in its main physical characteristics. The maximum values of matter density and pressure should also be associated at the centre of the configuration and should be monotonically decreasing functions of the radial coordinate towards its surface. These novel characteristics are required to explain some real objects such as white dwarfs, neutron stars, and even quark stars. In addition, there are additional components that are as important to the study of compact structures and that offer a more accurate picture of the behavior of celestial bodies. Anisotropies, for instance, might be present in the material composition  of the fluid sphere. In this context, anisotropy refers to the fact that the pressure in the radial direction and the pressure in the angular directions are not equal, or $p_r \neq p_t$. Therefore, $\Delta=p_t-p_r$ is used to define the anisotropy factor. 

\begin{figure}[htbp]
    \includegraphics[scale=.5]{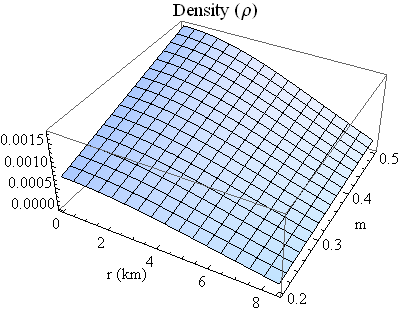}
        \includegraphics[scale=.5]{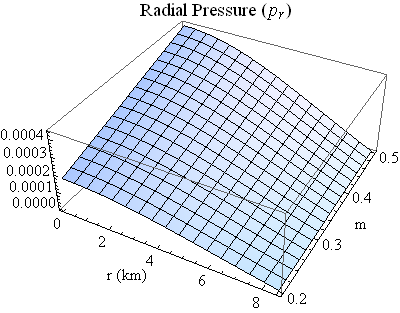}
        \includegraphics[scale=.5]{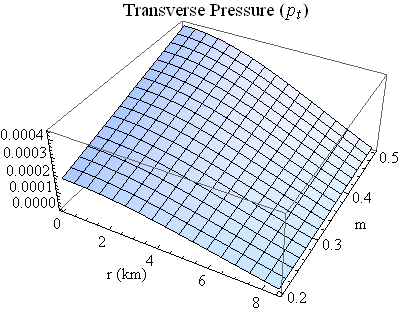}
          \includegraphics[scale=.5]{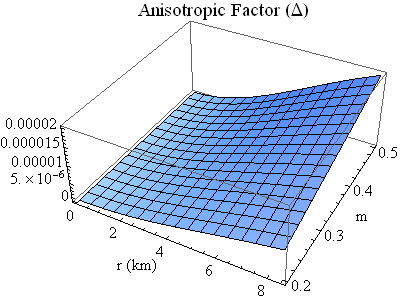}
       \caption{Matter density, radial pressure, transverse pressure, and anisotropic factor are shown against $r$\label{rho2}}
\end{figure}
\begin{figure}[htbp]
    \includegraphics[scale=.47]{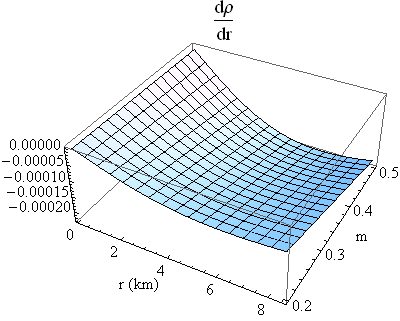}
        \includegraphics[scale=.47]{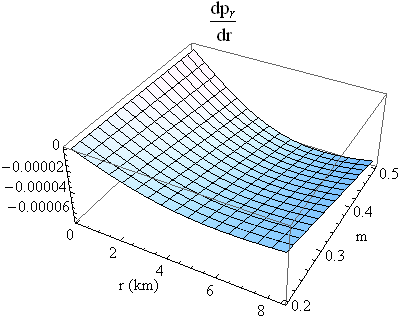}
        \includegraphics[scale=.47]{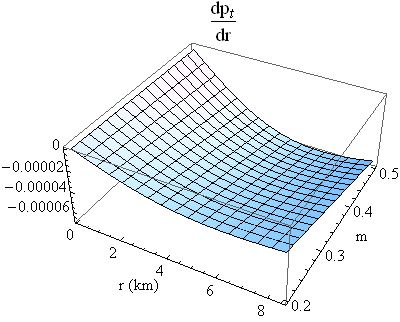}
       \caption{The density and pressure gradients are shown against `r'\label{grad2}}
\end{figure}

All thermodynamic observables $\rho,\,p_r$ and $p_t$ along with the anisotropy factor $\Delta$ are depicted in Fig.~\ref{rho2}. For a broad range of $m$, we may observe the behavior of matter density, radial pressure, and tangential pressure. It is important to see that these physical quantities monotonically decrease with increasing radial coordinates, with the highest values at the centre of the configuration. This graphic also depicts the behavior of the anisotropy factor $\Delta$. It behaves positively throughout the star, disappearing in the centre and increasing function of `r'.
The central values of density and pressure can be obtained as,
\begin{eqnarray}
\rho(r=0)&=&\frac{-3 a m+12 \pi  \beta +3 B m+16 \pi  B_g-n}{4 \pi  (3 \alpha -1)}, \\
p_t(r=0)&=&p_r(r=0)=\frac{3 a m \alpha - 3 B m \alpha + n \alpha - 
 16 B_g \pi\alpha - 4 \pi \beta}{4 \pi - 12 \pi \alpha}. 
\end{eqnarray}
The following two formulas will be utilized to determine the numerical values of the core density and central pressure for our current model, and they are shown in tabular form in our study.
Next, we are interested to find out the nature of the density and pressure gradients. Due to the complexity of the expressions of density and pressure gradients, we have taken the help of a graphical representation which has been shown in Fig.~\ref{grad2}. In the interior, all gradients had negative values, as depicted in the diagram.

\begin{table*}[t]
\centering
\caption{The numerical values of central density, surface density, central pressure, $\beta$, for the compact star Her X-1 for different values of `m' by taking $b=0.04 \times 10^{-5}$, n=0.005, $\alpha=0.3$}
\label{tb2}
\begin{tabular}{@{}cccccccccccccccc@{}}
\hline
m& $\rho_c$ & $\rho_s$ & $p_c$ &$\beta$  \\
& $gm./cm.^3$ &$gm./cm.^3$ & $dyne/cm.^2$\\
\hline
 0.2& $1.0016 \times 10^{15}$& $2.1428\times 10^{14}$ & $2.12577 \times 10^{35}$& 0.0000476416\\
 0.3& $1.44825 \times 10^{15}$ & $2.67264 \times 10^{14}$& $3.18866\times 10^{35}$&0.0000594217\\
 0.4& $1.89489 \times 10^{15}$ & $3.20248 \times 10^{14}$ & $4.25154 \times 10^{35}$&0.0000712018 \\
 0.5& $2.34154 \times 10^{15}$ & $3.73232\times 10^{14}$ & $5.31443\times 10^{35}$&0.0000829819 \\
\hline
\end{tabular}
\end{table*}

\subsection{Energy conditions}
\begin{figure}[htbp]
    \includegraphics[scale=.47]{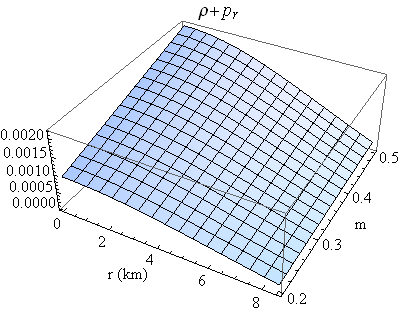}
        \includegraphics[scale=.47]{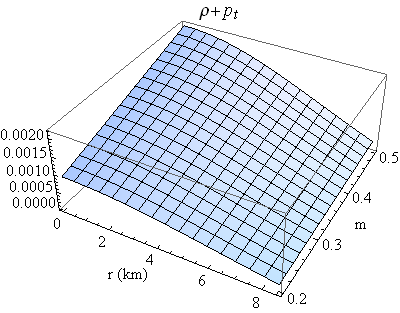}
        \includegraphics[scale=.47]{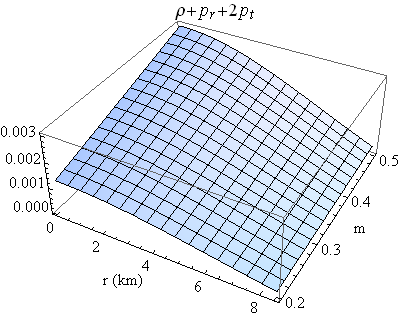}
        \includegraphics[scale=.47]{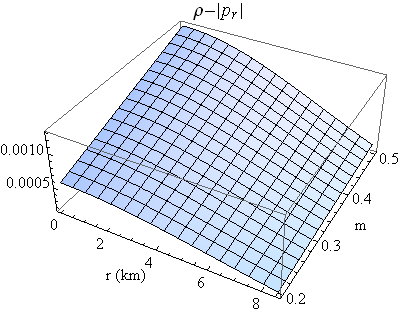}
        \includegraphics[scale=.47]{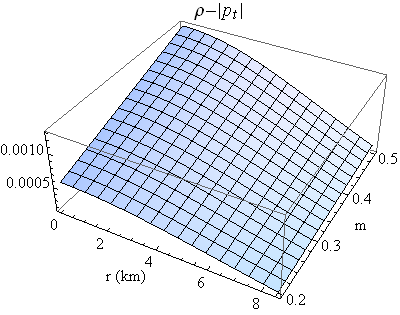}
       \caption{All the energy conditions are shown against `r'\label{en2}}
\end{figure}

All four of the energy conditions—the null energy condition (NEC), the weak energy condition (WEC), the strong energy condition (SEC), and the dominant energy condition (DEC)—are claimed to be met for a physically conceivable model if the parameters of the model, such as $\rho$, $p_r$, and $p_t$ satisfy the aforementioned expressions.
\begin{itemize}
    \item NEC: $\rho+p_r \geq 0,\, \rho+p_t \geq 0;$
     \item WEC: $\rho+p_r \geq 0,\, \rho+p_t \geq 0,\,\rho \geq 0;$
 \item SEC: $\rho+p_r \geq 0,\, \rho+p_t \geq 0,\,\rho+p_r+2p_t \geq 0;$
\item DEC: $\rho-p_r \geq 0,\, \rho-p_t \geq 0,\,\rho \geq 0$
\end{itemize}
It plays an essential role in comprehending the nature of matter as well \cite{Gasperini:2002bn}. In the context of GR, the wormhole model was considered as a way to explain how the energy criteria would be violated if exotic matter is present within the object. If these conditions are satisfied, it is shown that ordinary stuff exists. For $m \in [0.2,\,0.5]$, we graphically verified the validity of these conditions in Fig.~\ref{en2}, and we can observe that the previously stated energy requirements are all satisfied by the suggested hybrid star model in $f(Q)$ gravity.

\subsection{Equation of state}
Another crucial step is finding the equation of state, i.e., a link between pressure and density.
The radial pressure and matter density are assumed to be linearly related in the model by solving the field equations; however, the relationship between the transverse pressure and matter density is still uncertain. 
The equation of state parameters, usually denoted by $\omega_r$ and $\omega_t$, are two dimensionless quantities that can be used to characterize the relationship between matter density and pressure. For our current model, the equations of state parameters $\omega_r$ and $\omega_t$ are defined as follows:
\begin{eqnarray}
    \omega_r=\frac{p_r}{\rho},
    \omega_t=\frac{p_t}{\rho}.
\end{eqnarray}
\begin{figure}[htbp]
    \includegraphics[scale=.47]{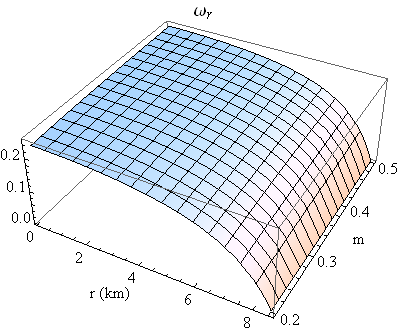}
        \includegraphics[scale=.47]{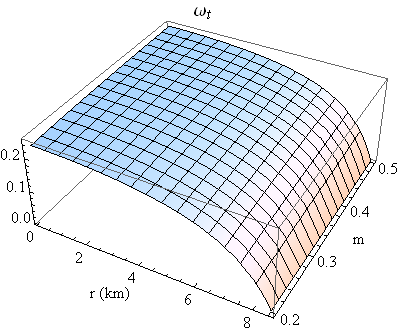}
       \caption{$\omega_r$ and $\omega_t$ are shown against `r'\label{omega2}}
\end{figure}
For a particular range of $m$, we have drawn the profiles of both $\omega_r$ and $\omega_t$ in Fig.~\ref{omega2}. The results clearly show that these two traits were most valuable near the star's center and decreased toward the edge. Furthermore, they fall inside the range of radiation era, i.e., $0<\omega_r,\omega_t<1$ \cite{Sharif:2016xbn}.

\section{Stability analysis of the present model}
In this part, we will examine the stability of our current model using (i) the causality condition, (ii) the adiabatic index, and (iii) the TOV equation which will be explained separately.
\subsection{Velocity of sound and cracking method}
\begin{figure}[htbp]
    \includegraphics[scale=.47]{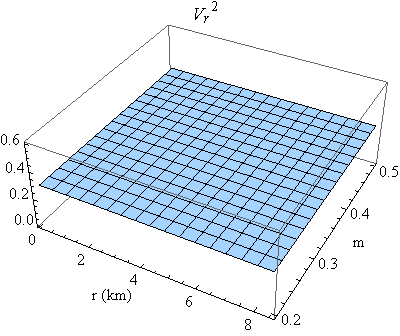}
        \includegraphics[scale=.47]{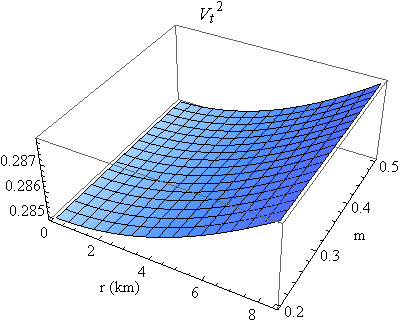}
        \includegraphics[scale=.47]{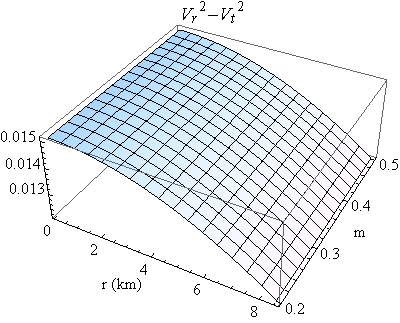}
       \caption{$V_r^2,\,V_t^2$ and $V_r^2-V_t^2$ are shown against `r'\label{sv2}}
\end{figure}

It is important to verify the causality requirement, which states that the speed of sound inside the compact object must be subluminal, in order to generate a physically accurate model.
 The following formula can be used to calculate a stellar fluid's sound speed.
 \begin{eqnarray}
     V_r^2=\frac{dp_r}{d\rho},
V_t^2=\frac{dp_t}{d\rho}.
\end{eqnarray}
We have chosen a linear equation of state between the radial pressure $p_r$ and the matter density $\rho$ for our current model. As a result, the speed of sound in the radial direction for our current model is simply set at $\alpha$ and does not vary on $m$. The tangential component, however, is dependent on the behavior of the anisotropy factor. Fig.~\ref{sv2} illustrates the variation of the square of the radial and transverse velocity, and it can be seen that the tangential velocity is increasing outward and less than $1$ for all values of $m$ throughout the star. As a result, we may assert that our model meets the causality constraint.\par
In a series of lectures \cite{55,56,57}, Herrera and colleagues in-depth examined the idea of cracking for stellar structures by taking into account anisotropic matter structures. The idea of cracking (or overturning) was first suggested in 1992. This method is beneficial for identifying potentially unstable anisotropic matter structures. They looked at the possibility of stability in the region of the star interior where the radial velocity of sound is greater than the transverse velocity of sound. We have generated the profile of $V_r^2-V_t^2$ in Fig.~\ref{sv2} to confirm this criterion, and the profile guarantees the potential stability of the current model.

\subsection{Adiabatic Index}
\begin{figure}[htbp]
    \includegraphics[scale=.47]{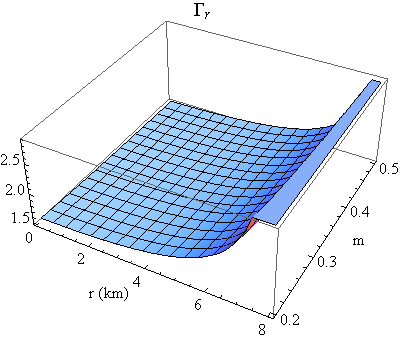}
        \includegraphics[scale=.47]{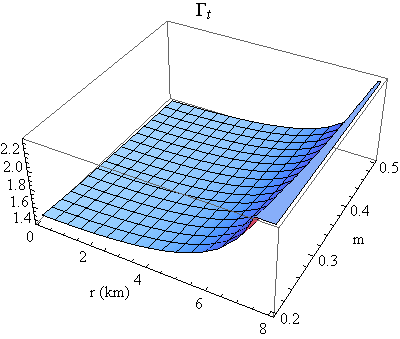}
       \caption{Relativistic adiabatic index $\Gamma_r$ and $\Gamma_t$ are shown against `r' \label{gamma7}}
\end{figure}

In this paragraph, we will analyze a crucial and important ratio of the two specific temperatures offered by $\Gamma$ in order to examine the area of stability of the hybrid star model. Chan et al. \cite{chan} proposed the concept of the adiabatic index for an isotropic fluid sphere, however, Chandrasekhar  \cite{Chandrasekhar:1964zz} was one of the first in this age to examine using the adiabatic index to look at the zone of stability for spherical stars. The expression for the adiabatic index changes as follows in the presence of pressure anisotropy:
\begin{eqnarray}
   \Gamma_r=\frac{\rho+p_r}{p_r}\frac{dp_r}{d\rho},\\
\Gamma_t=\frac{\rho+p_t}{p_t}\frac{dp_t}{d\rho}. 
\end{eqnarray}
The circumstances of stability are satisfied by the stellar object when the above two expressions take a value of more than $4/3$ according to Heintzmann and Hillebrandt's study \cite{hh}. Since it is impossible to verify this requirement analytically for the complexity of the expressions. We have drawn the profiles of $\Gamma_r$ and $\Gamma_t$ for various values in Fig.~\ref{gamma7}. The graphic shows that both $\Gamma_r$ and $\Gamma_t$ take values greater than $4/3$ across the fluid sphere, which ensures that the stability criterion is fully met.

 \subsection{The equilibrium under different forces}
\begin{figure}[htbp]
    \includegraphics[scale=.5]{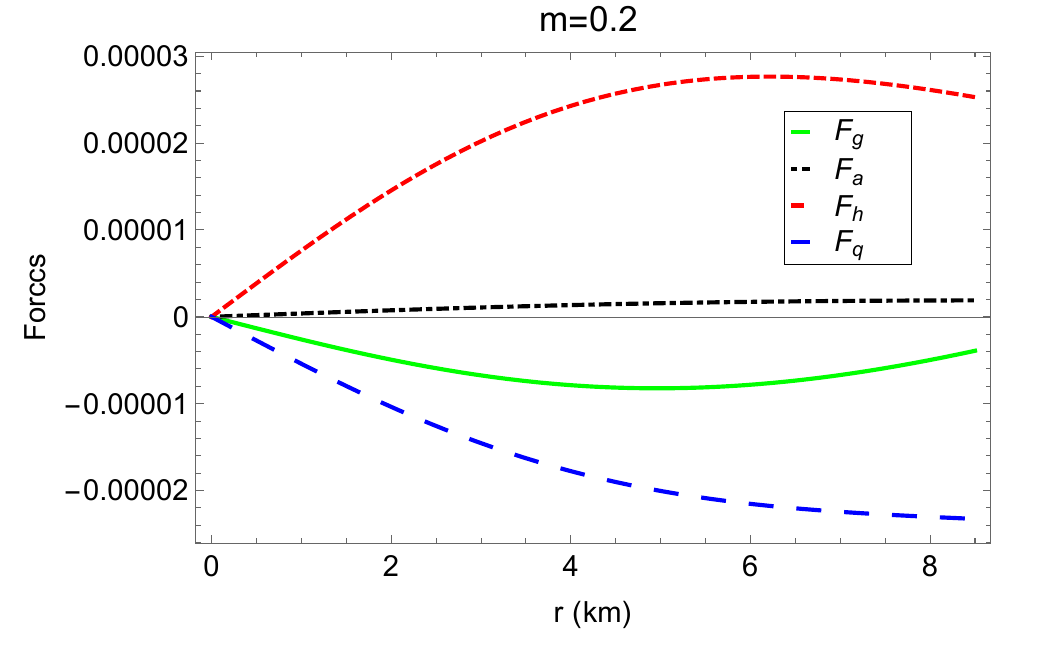}
        \includegraphics[scale=.5]{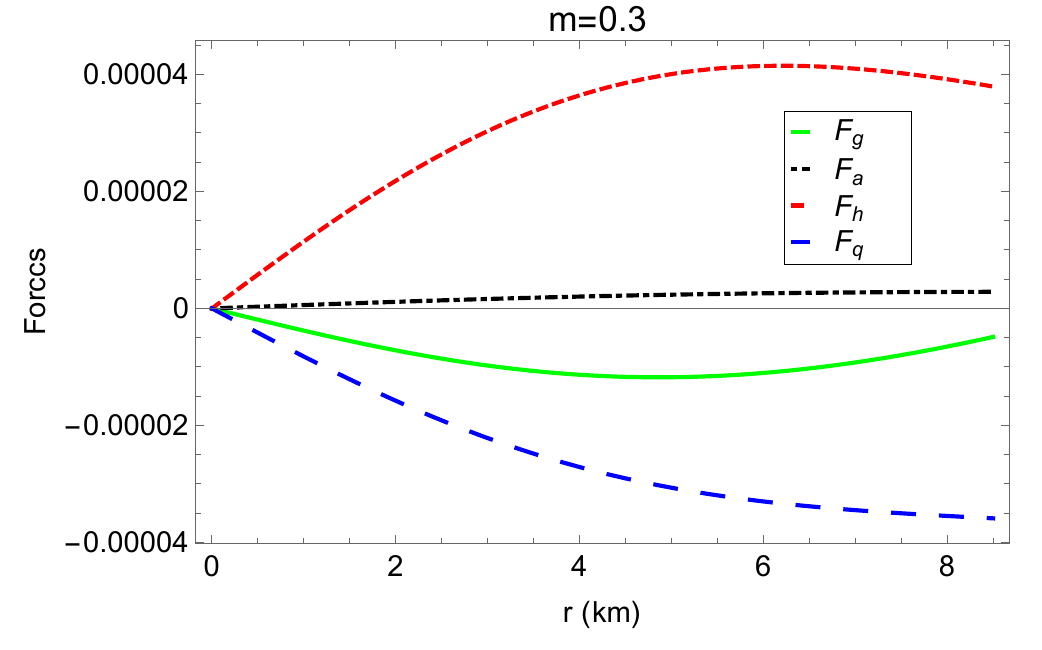}
        \includegraphics[scale=.5]{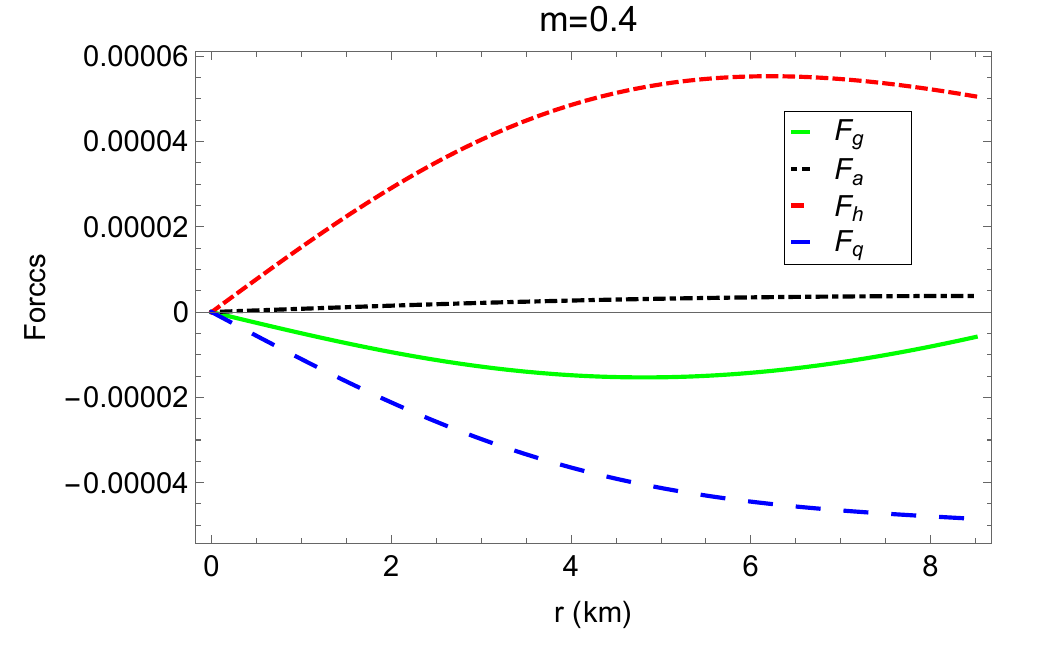}
          \includegraphics[scale=.5]{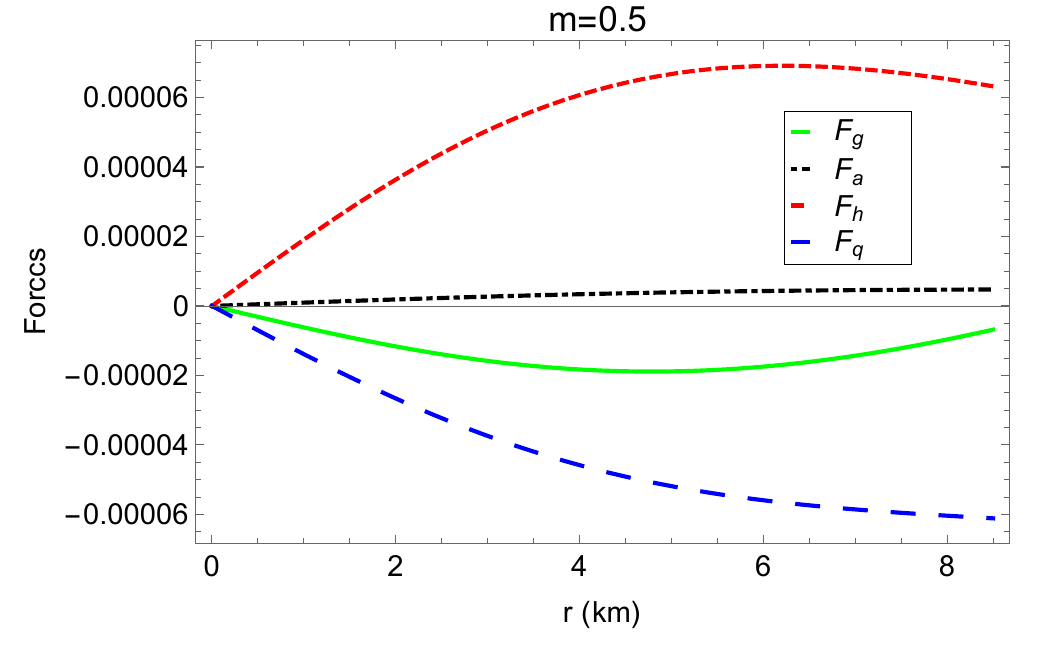}
       \caption{The different forces acting on the system are shown against $r$\label{tov5}}
\end{figure}

This subsection will examine the equilibrium of the model under various forces that are currently acting on the system. The four forces that constitute the equilibrium equation are the hydrostatic force ($F_h$), gravitational force ($F_g$), anisotropic force ($F_a$), and lastly the force associated with quark matter ($F_q$). Additionally, the explicit form of these forces is as follows:
\begin{eqnarray*}
F_g&=&-\frac{\nu'}{2}(\rho+p_r),\\
F_h&=&-\frac{dp_r}{dr},\\
F_a&=&\frac{2}{r}(p_t-p_r)=\frac{2}{r}\Delta \\
F_q&=&-\frac{\nu'}{2}(\rho_q+p_q)-\frac{d}{dr}(p_q),
\end{eqnarray*}
The Tolman–Oppenheimer–Volkoff (TOV) equation for our present model can be written as,
\begin{eqnarray}\label{con}
-\frac{\nu'}{2}(\rho+p_r)-\frac{dp_r}{dr}+\frac{2}{r}(p_t-p_r) -\frac{\nu'}{2}(\rho_q+p_q)-\frac{d}{dr}(p_q)=0,
\end{eqnarray}
Now the above equation can be denoted by,
\begin{eqnarray}
F_g+F_h+F_a+F_q &=& 0.
\end{eqnarray}
Fig.~\ref{tov5} shows the formulation of various forces acting on our system for different values of the coupling parameter $m$. From the figure, we can see that the combined effects of all four different forces make our model stable. 

\section{Discussion}
In the present work, we propose a model of a hybrid star in the realm of $f(Q)$ modified gravity. We have chosen the Tolman-Kuchowicz metric potential to solve the field equations. The obtained model has been matched successfully to the exterior spacetime. The most significant findings include the following: Our results show that the energy density $\rho$, pressures $p_r$, $p_t$, of the investigated compact star approach their greatest value near the core, while they are at their minimum at the surface. It is crucial to note that the radial pressure $p_r$ at the surface of the star vanishes. The central density $rho_c$ approaches a significantly enormous value when we are dealing with the core of the star, and it makes the stars very compact. The high compactness offers a proper justification for the validation of the $f(Q)$-model that we propose. The numerical values of central density, surface density, and central pressure have been calculated for various values of $m$, and it is clear that as $m$ rises, all three variables take on increasing values. At the same time, the $\beta$ increases as $m$ grows.
The relevance of the surface redshift is increased by the existence of anisotropies in the stellar content, which improves the stability and balancing processes. The contribution that it will make to the equilibrium mechanism, however, relies on the sign, or whether it is positive or negative, according to $p_t>p_r$ or $p_t<p_r$. In the first scenario, the system experiences a repulsive force that reduces the gravitational gradient, whereas in the second scenario, the force conveyed by anisotropy contributes to the gravitational force compressing the star. The structure will eventually keep collapsing till its Schwarzschild radius if the pressure of nuclear force is insufficient to push against gravity. The object then generates a black hole with a variety of peculiar characteristics. This indicates that the equilibrium and stability of the configuration are affected by the presence of an attracting force caused by anisotropies. We developed a graphical diagram to illustrate the anisotropic behavior. The anisotropic force shown in Fig~\ref{rho2} is repulsive in nature for our present model.\par
Taking into account the hybrid star, we also found that a number of energy conditions are satisfied, which further shows that there is no exotic matter present and that the underlying matter distribution is completely non-exotic matter. It should be noted that stability analysis is crucial for modeling any compact object. The causality requirement is met by the current model. In this case, stability is investigated using cracking methods. Our recommended models are conceivably reliable against the variations, according to the stability study proposed by Herrera. The relativistic adiabatic indices $\Gamma_r$ and $\Gamma_t$ are shown, and they both assume values greater than $4/3$, satisfying the stability requirement. Two different EoS parameters, $\omega_r$ and $\omega_t$, are involved in the anisotropy investigation. The range of realistic and normal distribution of matter is determined by these two Eos parameters. The maximum allowable mass and the corresponding radius are obtained and it relates to the mass of compact stars found in the literature. Another crucial point is that the measurements of mass and bag constant $B_g$ have been studied in detail via contour plots. From our analysis, we have obtained the range of bag constant $B_g$ as $55-95~MeV/fm^{-3}$ which is very much compatible with CERN data about quark-gluon plasma (QGP) as well as compatible with the RHIC preliminary results \cite{h1,h2} and the observational result by  Farhi and Jaffe \cite{farhi}.\par 

Many stellar solutions has been obtained in $f(R)$, $f(R,T)$ gravity, etc. to verify the reliability of these types of modified gravity. These types of gravity are based on the Riemannian geometry, where torsion and nonmetricity are zero. Within this framework, the Ricci scalar curvature works as a building block of space-time. But here, we represent the work to see the behavior of the stellar model when the gravitational interaction between two particles in space-time is described by the nonmetricity $Q$, upon which $f(Q)$ gravity theory is established. We have used $f(Q)$ gravity to verify whether it gives the same physical properties of the stellar model as the previous result, like realistic gravity. There are a number of works on compact stars in the framework of Einstein's GR as well as in modified gravity. To compare our results with those types of realistic gravity like $f(R)$, $f(R,T)$ gravity etc. one can see the references \cite{GGL, SD, G}.
The success of our recommended model was confirmed throughout the study in conjunction with a proper contrast of a large number of compact star candidates. As a result, the implications of our chosen methodologies provide a better justification for compact objects. As a result, we draw the conclusion that our suggested hybrid star model behaves successfully and adequately explains the physical characteristics in the circumstances of $f(Q)$ gravity.

\section*{Acknowledgements} P.B. is thankful to the Inter-University Centre for Astronomy and Astrophysics (IUCAA), Pune, Government of India, for providing visiting associateship. PB also acknowledges that this work is carried out under the research project Memo No: $649$(Sanc.)/STBT-$11012(26)/23/2019$-ST SEC funded by the Department of Higher Education, Science \& Technology and Bio-Technology, Government  of West Bengal. SP \& PKS  acknowledges the National Board for Higher Mathematics (NBHM) under the Department of Atomic Energy (DAE), Govt. of India for financial support to carry out the Research project No.: 02011/3/2022 NBHM(R.P.)/R \& D II/2152 Dt.14.02.2022. A. Malik acknowledges the Grant No. YS304023912 to support his Postdoctoral Fellowship at Zhejiang Normal University, China.

\end{document}